# Selection and Characterization of Commercial Precursors, GIC-based, for Industrial Production of Bulk Graphene Nanoplatelets

Francesco Cristiano[1], Francesco Bertocchi[1], Mohab Elmarakbi[2], Ahmed Elmasry[2] and Ahmed Elmarakbi[2,*]

[1]*Nanesa srl, Via Del Gavardello 59/c 52100 Arezzo (AR), Italy*

[2]*School of Engineering, Faculty of Technology, University of Sunderland, Sunderland SR6 0DD*

[3]*Department of Mechanical and Construction Engineering, Faculty of Engineering and Environment, Northumbria University, Newcastle NE18ST, United Kingdom*

*Corresponding author: Ahmed Elmarakbi (ahmed.elmarakbi@northumbria.ac.uk)

**ABSTRACT**

The morphology and the chemical characteristics of graphene nanoplatelets are important parameters to define the potential of these particles in various applications. In this paper we firstly conducted a market analysis to identify commercial CIG (Graphite Intercalation Compounds) with different characteristics, our selection was based on physical-chemical criteria (such as purity, mesh, expansion degree) and commercial (cost, availability, etc). The materials were prior expanded and exfoliated on a laboratory scale, then on an industrial pilot plant, to study the final characteristics of the graphene nanoplatelets and the relations of them to the starting materials. Selected materials and products, derived from the exfoliation process, have been well characterize; SEM, OM, XRD, PSA, BET for the morphological characteristics and TGA, FT-IR, XRF, EDS for chemical ones. We have obtained particles with different chemical-physical characteristics, potentially suitable for applications that request to improve electrical and thermal conductibility and/or mechanical reinforcement and/or barrier effect.

# 1. INTRODUCTION

The choice of the starting material is of fundamental importance for the quality and the final characteristics of the product. In the market there are many types of graphite based products. Graphite is a raw material with a unique blend of physical and chemical properties [1-2] . There are quite a number of minerals similar in appearance to graphite; however, graphite's intrinsic properties make it easy to distinguish. A useful classification of graphite depends on the mode of formation that leads to three physically distinct common varieties:

- *amorphous (micro-crystalline) graphite, which has a carbon content of 70-85%;*
- *high crystalline graphite (lump, vein or crystalline vein), which has a carbon content of 90-99%;*
- *flake graphite, which has a carbon range of 80-98%. Flake graphite (i.e., flat plate-like grains from <1 mm to 2.5 cm in size) is sold in two particle size distributions: coarse flake (-20 to +100 mesh) and fine flake (-100 to +325 mesh).*

Synthetic graphite has a higher purity but lower crystallinity than natural graphite and is divided into: primary or electrographite, with a carbon content of 99.9%, which is manufactured on a large scale in electric furnaces using calcined petroleum coke and coal tar pitch (used to produce electrodes and carbon brushes); secondary synthetic graphite in the form of powder or scrap, which is produced by heating calcined petroleum pitch (used in the refractories industry); and graphite fibres, which are produced from organic precursors such as rayon or polyacrylonitrile and tar pitch (used as reinforcing agents in polymer composites in aerospace and sporting goods).

There are significant differences between natural and synthetic graphite, natural graphite is generally less pure than its synthetic equivalent [4]. The synthetic graphite is less conductive due to a more defective crystal structure. Natural graphite usually has to be purified and upgraded, synthetic graphite can be engineered that has 99.9% and higher carbon. Technology now allows natural graphite material to be upgraded to more than 99.5% carbon. Purification techniques have improved to the point that even low-quality graphite can be used in high-tech applications that were once the domain of synthetic material.

The intercalated graphite is obtained by chemical intercalation of graphite flake with sulfuric acid and nitric acid. If subjected to a strong heat source, the intercalation compounds vaporize immediately creating strong pressures between the graphite layers by changing the structure from flake to worm-like. The approach will start from low costs materials such as commercial GIC or different kind of Graphite, with selected mesh size and purity, treated with a low environmental impact and high efficiency continuous technologies.

The intercalation compounds of graphite are interstitial compounds in which the foreign species is included in the interplanar interstitial sites of the graphite crystal such that the layer structure of the graphite lattice is retained [5-6]. These compounds are the most well-known of all the compounds of graphite. Graphite reacts with a large number of acids to form intercalation compounds which have been referred to as "acid salts of graphite". These acids act as electron acceptors in the graphite crystal by forming negatively charged acids radicals ($NO^-_3$, $HSO^-_4$, etc.). However, only a fraction of the acid molecules undergoes this ionization. The rest remains as acid molecules in the graphite crystal. Graphite-$H_2SO_4$ (also known as graphite bisulfate or expandable graphite) is the most extensively studied compound in this category. Graphite bisulfate consists of graphite layers with $HSO_4$-ions and $H_2SO_4$ molecules between the layers [7-8]. The blue stage 1 graphite bisulfate lamellar compound can be prepared by direct chemical interaction of graphite with a mixture of concentrated sulfuric acid and an oxidizing agent (nitric acid, chromic oxide, potassium permanganate, ammonium persulfate, manganese dioxide, lead dioxide, arsenic pentoxide, iodic acid, periodic acid or manganese salts) or by electrolysis [9-10].

## 2. METHOD

SEM analysis were performed with a Scanning Electron Microscope *EVO MA10 Zeiss* in Secondary Electron Imaging, for EDS analysis was used *Back Scattered Electrons (BSE)* mode with INCA software. For the analysis, GIC's samples was placed on Carbon Conductive Tabs, while the graphene nanoplatelets were placed on a copper foil. Copper was especially useful for the EDS analysis, in order to distinguish the elements during the scan in BSE mode. The Thermogravimetric analysis (TGA) was performed with *SDT Q600 TA*. The analysis were conducted on GIC Samples from 25 °C to 1000 °C in air with a heating ramp of 10°C/min.

For the analysis was used the quantity of about 8-10 milligrams for each GIC sample. For XRF analysis was used Sequential XRF *Thermo ARL Advant'x.* They were analyzed GIC samples and Graphene nanaplatelet dried powders. Particle Size Analysis (PSA) was performed with *HELOS Sympatec* with *Sucell* dispersion system. PSA was used for graphene nanoplatelet powders. They have been dispersed in Ethanol with Sucell ultrasound system. A *LEICA Microscope DM2500MH* was for Optical Microscopy in reflection mode. It was used to investigate on graphene nanoplatelets. Samples were dispersed in Acetone with a mild sonication. A drop of the dispersion was placed on a laboratory glass. OM was performed after the evaporation of the solvent.

FT-IR analysis was performed with a *PERKIN-ELMER MOD. GX.* GIC samples and Graphene Oxide were studied. The analysis were conducted in transmission mode on KBr/Graphene pellets. The pellets were obtained by grinding and compressing (under vacuum) the dried powders of potassium Bromide and Graphene. The concentration of Graphene was 1%. The viscosity of the dispersion during the exfoliation process was monitored every hour with a *Thermo Scientific Haake Viscotester 1 Plus.* The XRD analysis was performed in collaboration with Prof. Gaetano Guerra, Dep. Chemistry and Biology, University of Salerno. TEM and HR-TEM analysis were performed in collaboration with Institute for Polymers, Composites and Biomaterials (IPCB), Portici (Naples).

**2.1 Materials**

*2.1.1. Raw Materials- precursors for Graphene derivative*

The Expandable Graphite (GIC) was supplied from: Asbury Carbons, GK-Graphite, Faima, Luh, Nyacol. Have been used the following codes: IG1, IG2 (Asbury); IG5 (GK), IG6 (Faima), IG7 (Luh), IG11 (Sanyo).

*2.1.2. Selection Criteria*

The choice of precursor was based on two criteria: chemical-physical criteria (purity, mesh size, potential expansion rate), commercial criteria (cost, availability, also on large quantities). The purity is a parameter related to the carbon content compared to any impurities such as metals and silica. The mesh size indicates the particle size of the graphite flakes. Expandable graphites with very different particle sizes could lead to very different products. The expansion rate indicates the potential Volume increase obtained after the expansion phase. This parameter was very important to get indirect information about the degree of intercalation of graphite and the morphology of the expanded.

The cost of materials was also an important parameter, for a production on an industrial scale was essential to use not expensive materials. The reference threshold was up to 20 euro per kilogram. It was also important to check the availability of the supplier to ensure high amounts of material and guarantee the quality standards. In Table 1, the list of precursors chosen with these criteria is presented.

*Table 1:* List of precursors

| ID | Type | Supplier | Nominal Size ($\mu m$) | Carbon (%) | Expansion Ratio (cc/g) |
|---|---|---|---|---|---|
| NNSa IG1 | Expandable Graphite | Asbury | 180 | 90 | 215 |
| NNSa IG2 | Expandable Micro-Graphite | Asbury | 75 | 80 | 30 |
| NNSa IG5 | Expandable Graphite | GK | 300 | 98 | 350 |
| NNSa IG6 | Expandable Graphite | Faima | 300 | 95 | 250 |
| NNSa IG7 | Expandable Graphite | Luh | 250 | 90 | 250 |
| NNSa IG11 | Expandable Graphite | Nyacol | 300 | 99 | 350 |

## 2.2 Process, development and optimisation

To study the precursors reported in Table 1, the work was divided in two parts. In the first part was performed a preliminary study on a laboratory scale of all selected precursors, performing a complete characterization, in order to evaluate all properties of precursors. Subsequently, materials were processed with microwave and ultrasounds to exfoliate the graphite nanoplatelets. The products were then characterised with various analysis in order to define all properties. In the second step, industrial scale, the precursors selected in the preliminary study were processed on an industrial plant for the exfoliation of graphite nanoplatelets. The products obtained were analyzed and the properties were compared with laboratories ones.

## 2.3 Exfoliation and Analysis

### 2.3.1 Lab-scale process

This process consisted in two steps: *i) Expansion with microwaves; and ii) Exfoliation with ultra-sonication*

i) <u>Expansion</u>

The expansion phase was performed using a microwave system (Fig.1). On heating, the graphite flakes spontaneously expand and increase their volume several hundredfold in comparison to their original volume. Typically a worm-like structure is generated from each flake. This system, discontinuous, guarantees a considerable thermal shock that causes immediate expansion of graphite. For the screening trials a simple set-up was chosen consisting of a large gastronomic microwave oven working at 2.45 GHz with 4.8 kW maximum power output.

A small amount of each sample (2g of powder) was placed as thin film in a quartz glass vessel the centre of the oven and heated there typically at 2kW for 12s. The aim was to expand 100% of the material without generating unrequested plasma sparks. These heating parameters seemed to be ideal for all 6 samples (IG1-IG11). All samples showed a heavy expansion reaction, starting to glow, sparkle and expand after 4 seconds of heating - and after 12 seconds the complete sample was expanded to 0.6 to 0.8l of volume and the process finished.

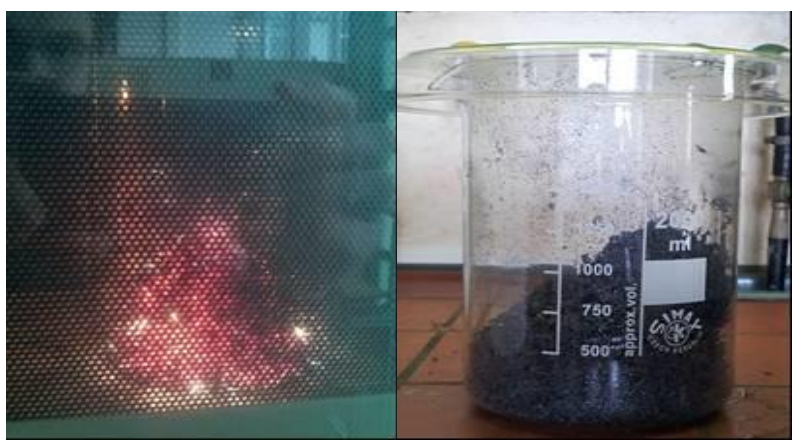

***Figure 1.*** *Left: heavy expansion reaction while heating; Right: expanded sample after heating*

This was a positive intermediate result: each proposed precursor material showed good processability with microwave heating. Afterwards, 10g of each material were expanded in that way for a subsequent exfoliation process step with ultra-sonification.

ii) <u>Exfoliation</u>

For this preliminary stage was used a 1000 Watt tip sonicator. The sonicator is an ultrasound device developed for laboratory tests and for the treatment of industrial liquids (Fig. 2). It is composed by a current generator connected to a transducer that converts electrical energy into mechanical oscillations. These oscillations propagate in the form of ultrasound through metal structures (horns, boosters) coming to the liquid to be treated. With special mechanical devices (geometry of the sonotrode and the booster), it is possible to vary the amplitude of the output and it is also possible tune the output power from 50 to 100%. Some technical characteristics of the instrument, includes: Efficiency> 85; Operating frequency: 20 kHz; and Maximum power: 1000 W.

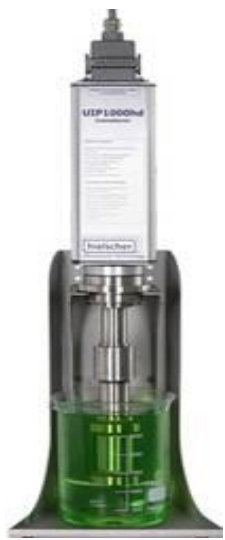

*Figure 2. Configuration for ultrasound system used for the exfoliation of graphite nanoplatelets in laboratory-scale process.*

For this work was used a discontinuous configuration with the horn partially immersed in the liquid, stirred by a magnetic anchor. The parameters, amplitude and power, have been calibrated with those of plant for the continuous process. As solvent was used demineralized water (about 800ml) in which the expanded graphites were pre-mixed.

The parameters were set for all tests as: *Quantity of expanded: 3 grams; Amount of water: 800ml; Ambient temperature; Configuration: Intensive treatment; Sonification time: 60 minutes; and Pre-mixing time: 5 minutes.* After the exfoliation process (Fig. 3), dispersions were filtered. The paste recovered from the filter was dried in a ventilated oven at 90 ° C for 8 hours. The solid was then recovered and pulverized by stirring in a closed container. Below pictures of the different powders obtained (Fig. 4).

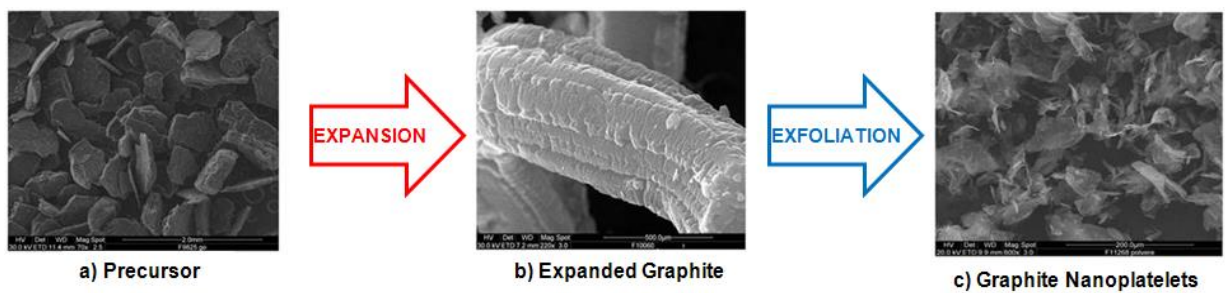

*Figure 3.* SEM images show the morphological change of the material during the various stages of the exfoliation process: a) Starting precursor in flake; b) Expanded worm-like structure of precursor after microwave expansion (expanded graphite); c) Graphite nanoplatelets obtained after exfoliation treatment with ultrasound sonication.

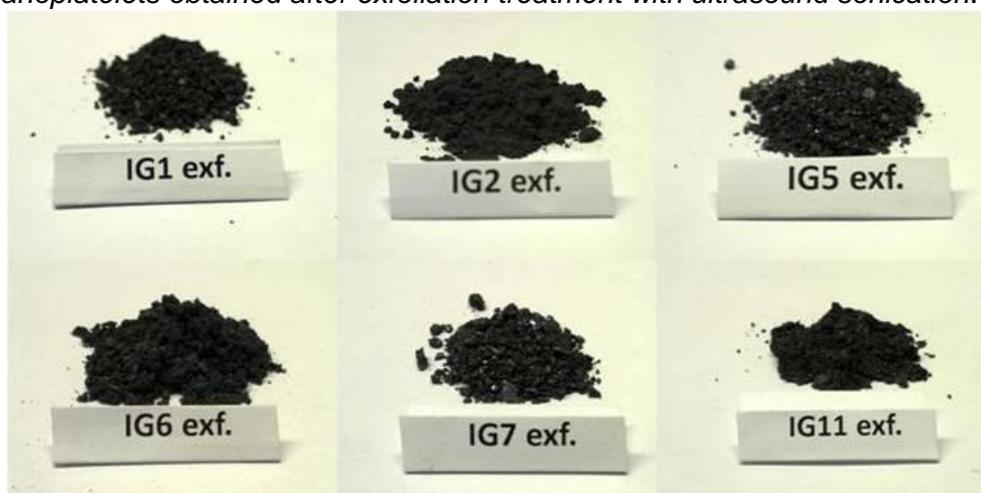

*Figure 4.* Pictures of powders obtained from different precursors. The procedure of exfoliation used was the same for all samples.

### 2.3.2 Industrial process

The industrial process (Fig. 5) involves a phase of expansion of precursors using a continuous-flow system of air or nitrogen heated at high temperature (850 °C). The hot gas, not only instantly expands the material but also acts as a transport for the same. The expanded graphite is cooled and classified (particle size) using separation systems and then exfoliated using ultrasounds. This takes place in the exfoliation plant, a continuous system composed of premixing zones of expanded graphite and solvent (typically water), a sonication chain composed of different sonicators arranged in series, a collector system to recover material (exfoliated) and separate from solvent. The method offers the advantage of a system composed preferably of three combined stages: a "pre-treatment" of the material through a system of volumetric sonication, a subsequent intensive treatment with a system of tip sonication and a final finishing treatment.

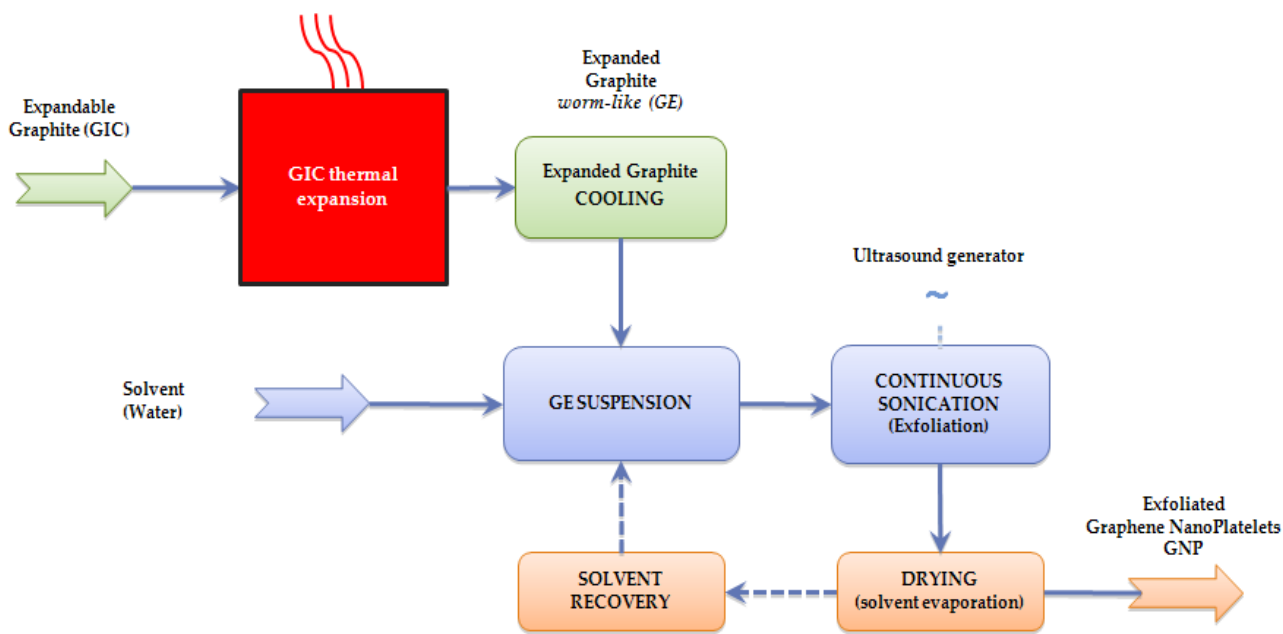

**Figure 5.** Block diagram of the continuous-flow industrial process.

Choice of the precursors for industrial process

Table1 presents a summary of all analysis performed on the precursors IG1-IG11, both before and after the exfoliation process. The last column shows the reasons for the choice of the precursors for the industrial test.

*Table 1.* Analysis summary

| ID | Declared Characteristics | Preliminary Analysis | Product Analysis (Lab-scale) | Choice for the industrial test |
|---|---|---|---|---|
| IG1 | Expandable graphite with mean nominal size | SEM: Good separation TGA: Good Weight loss XRF: High impurities content | PSA: Good VMD, Gaussian SEM: Good flatness, Good thickness, Regular Edges XRD: High Degree of stacking XRF: Regular Impurities content | NO high impurities content and High Degree of Stacking |
| IG2 | Expandable micro-graphite with low nominal size | SEM: Cross section not visible TGA: Low Weight loss XRF: High impurities content | PSA: HIgh VMD, Gaussian SEM: Good flatness, High thickness, Regular Edges XRD: Good Degree of Stacking XRF: Still High impurities content | YES only micro-graphite readily available on the market. |

| | | SEM: Good | | NO |
|---|---|---|---|---|
| IG5 | Expandable graphite with high nominal size and expansion ratio | SEM: Good separation TGA: Good weight loss XRF: Good impurities content | PSA: Good VMD, Gaussian SEM: Good flatness, Good thickness, Irregular Edges XRD: Low Degree of Stacking XRF: Regular Impurities content | NO Good properties but not particularly interesting compared with IG6 and IG11 |
| IG6 | Expandable graphite with high nominal size and good expansion ratio | SEM: High separation TGA: High Weight loss XRF: Good impurities content | PSA: Good VMD, Gaussian SEM: High flatness, Low thickness, Regular Edges XRD: Good Degree of Stacking XRF: Regular Impurities content | YES Good results in preliminary and lab-scal process. Very good morphology of platelets |
| IG7 | Expandable graphite with mean nominal size and expansion ratio | SEM: Good separation TGA: Good Weight loss XRF: Good impurities content | PSA: Good VMD, Gaussian SEM: Good flatness, Good thickness, Irregular Edges XRD: Good Degree of Stacking XRF: Regular Impurities content | NO for the same reasons of IG5 |
| IG11 | Expandable graphite with high nominal size and expansion ratio. High Carbon content (> 99%) | SEM: High separation TGA: High Weight loss XRF: Good impurities content | PSA: Low VMD, Gaussian SEM: High flatness, Low thickness, Regular Edges XRD: Good Degree of Stacking XRF: Regular Impurities content | YES Good results in preliminary and lab-scal process. Very good morphology of platelets |

According to the Table 1, the expandable graphites choices for testing on industrial plant were NNSa IG6, NNSa IG11 and NNSa IG2. All plant parameters have been optimized in order to maximize the ability of expansion and exfoliation of the materials.

### *NNSa IG6*

This expandable graphite presents the greatest weight loss after the expansion phase (about 40%). For this reason have been used 16 kg of material to obtain about 10 kg total of final product. The continuous expansion system worked in about 3 hours all 16 kg of graphite. The mean temperature generated by the flames, in the expansion phase, is approximately 900 °C. This system determines a very strong thermal shock that causes it to expand immediately Graphite without damaging because the residence time is very short (a few seconds). To wet the 10 kg of expanded graphite, were used 1000 liters of demineralized water (1% concentration). The expanded graphite was then premixed with water before starting exfoliation process, to ensure to all the expanded material to receive the same treatment. For the exfoliation phase was used a configuration for intensive treatment using mainly tip sonicators and system pressure to 2-3 bar. The system has worked in recirculation for about 6 hours at a power of 3000 Watt. The graph (Fig.6) shows the trend of the viscosity and temperature over time. It is observed an increase of the viscosity, according to a sigmoid trend, especially in the range between 120 and 300 minutes. After 300 minutes the viscosity tends to stabilize according to a steady state trend, this implies fairly that the maximum capacity of exfoliation is reached.

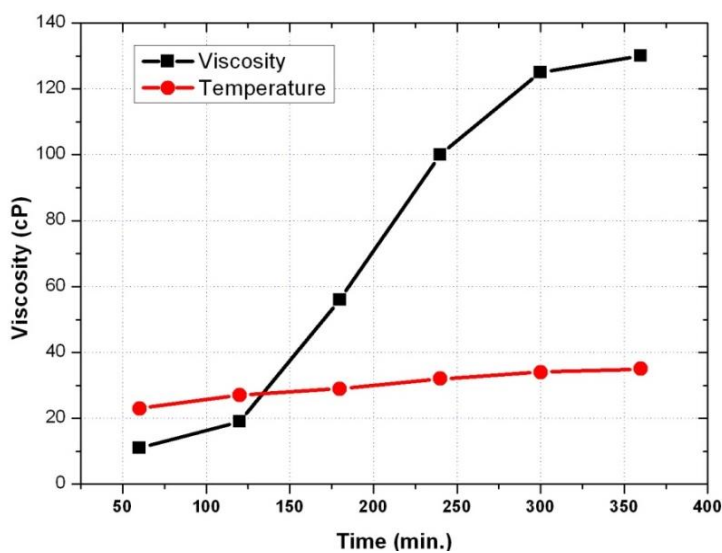

*Figure 6. The temperature and viscosity trend, versus the time process, of suspension (water/IG6 at 1% by wt).*

### *NNSa IG2*

For the micro-expandable graphite IG2 was used a similar procedure to that of graphite IG6. Were used 10 kg of expandable graphite (loss of weight of 9-10%), obtaining about 9 kg of finished product. The expansion and sonication parameters were set as before (IG6), to reach the maximum degree of expansion and exfoliation. To wet the micro-expanded graphite have been used 900 liters of water. The exfoliation process lasted a total of 10 hours. The increase in time of exfoliation was linked with slow and constant increase of viscosity as reported in the graph (Fig. 7).

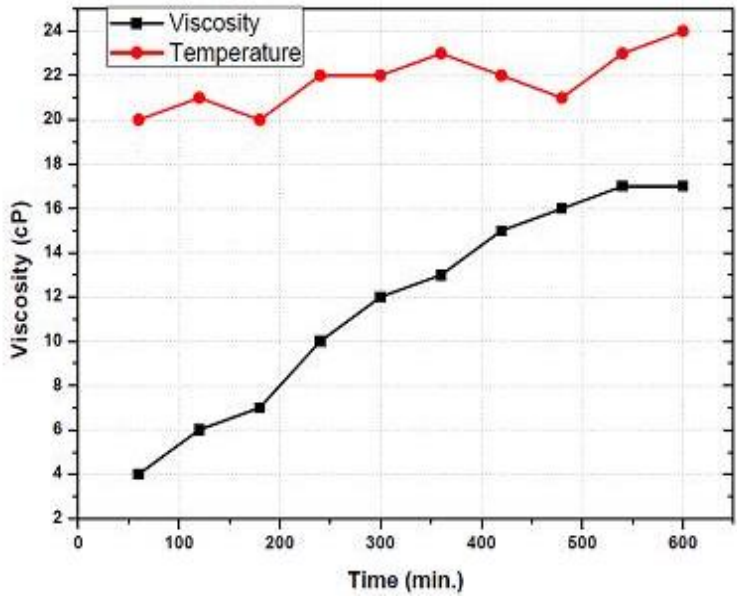

*Figure 7.* *The temperature and viscosity trend, versus the time process, of suspension (water/IG2 at 1% by wt).*

### *NNSa IG11*

Even for the Expandable graphite IG11 was used a similar procedure to that of graphite IG6. Were used 14 kg of expandable graphite (loss of weight of 30%), obtaining about 10 kg of finished product. To wet the micro-expanded graphite have been used 1000 liters of water. Using the same parameters of the previous trials, the exfoliation process lasted a total of 8 hours. In Fig. 8 the graph shows the viscosity trend in time.

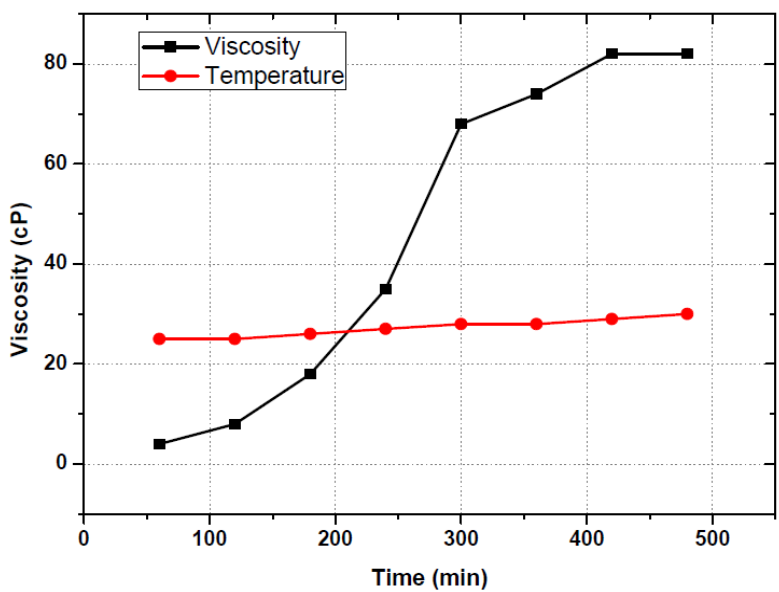

*Figure 8.* *The temperature and viscosity trend, versus the time process, of suspension (water/IG11 at 1% by wt).*

## 2.4 Measurements and Characterisation

### *2.4.1. Structural characterisation of precursors*

#### <u>*SEM*</u>

SEM analysis (Figs. 9-14) were performed to determine the morphology of six types of expandable graphite (NNSa IG1, NNSa IG2, NNSa IG5, NNSa IG6, NNSa IG7, NNSa IG11).

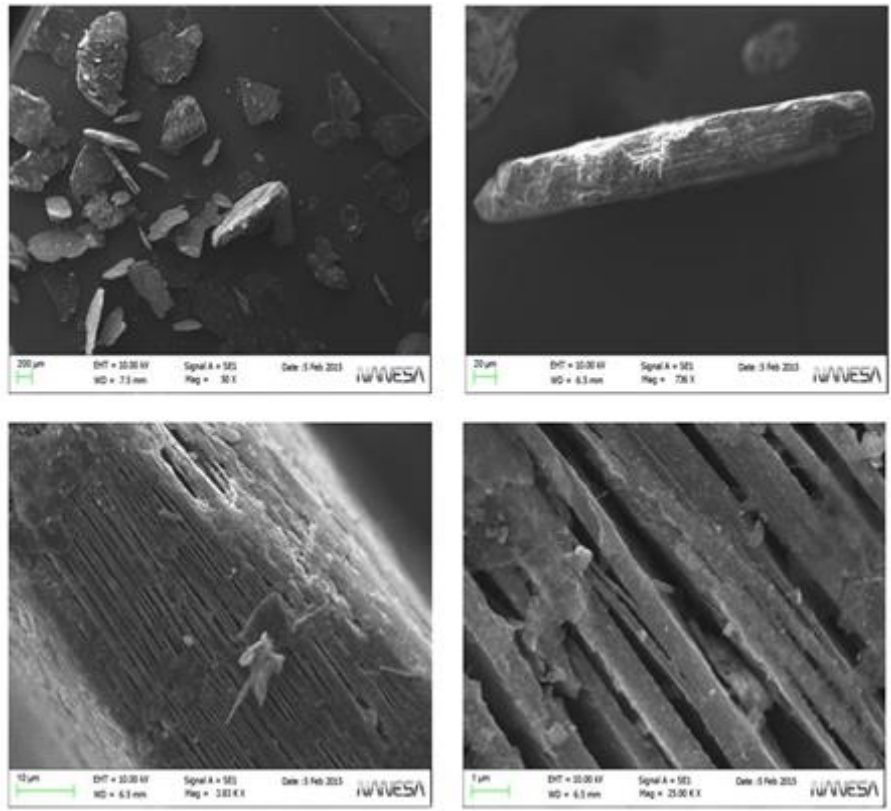

*Figure 9.* SEM images, Sample NNSa IG1 at various magnifications

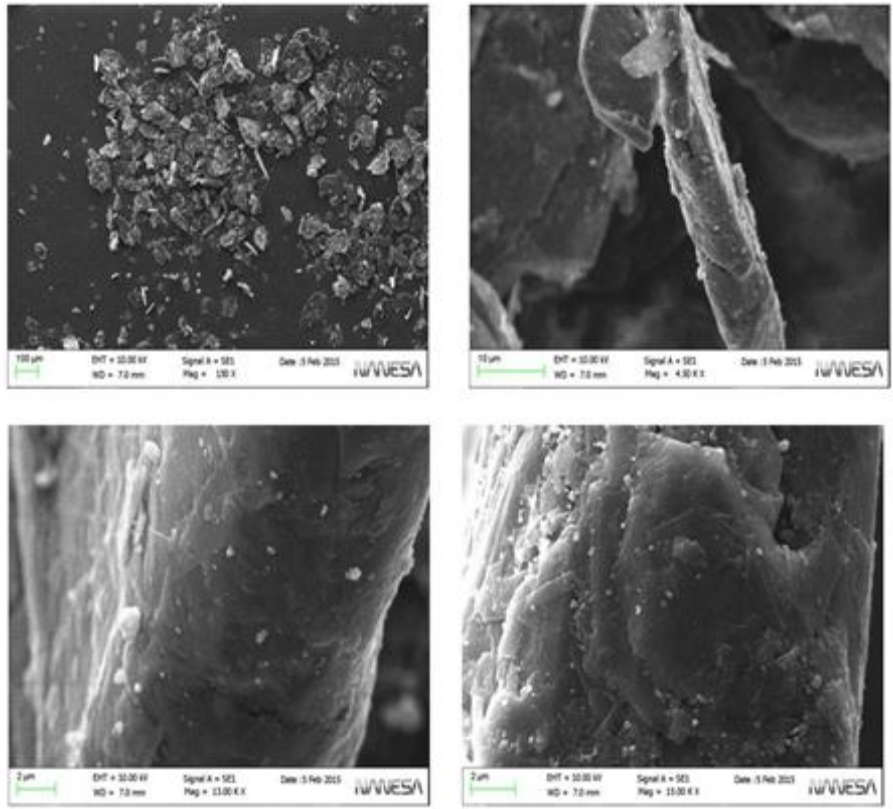

***Figure 10.*** *SEM images, Sample NNSa IG2 at various magnifications.*

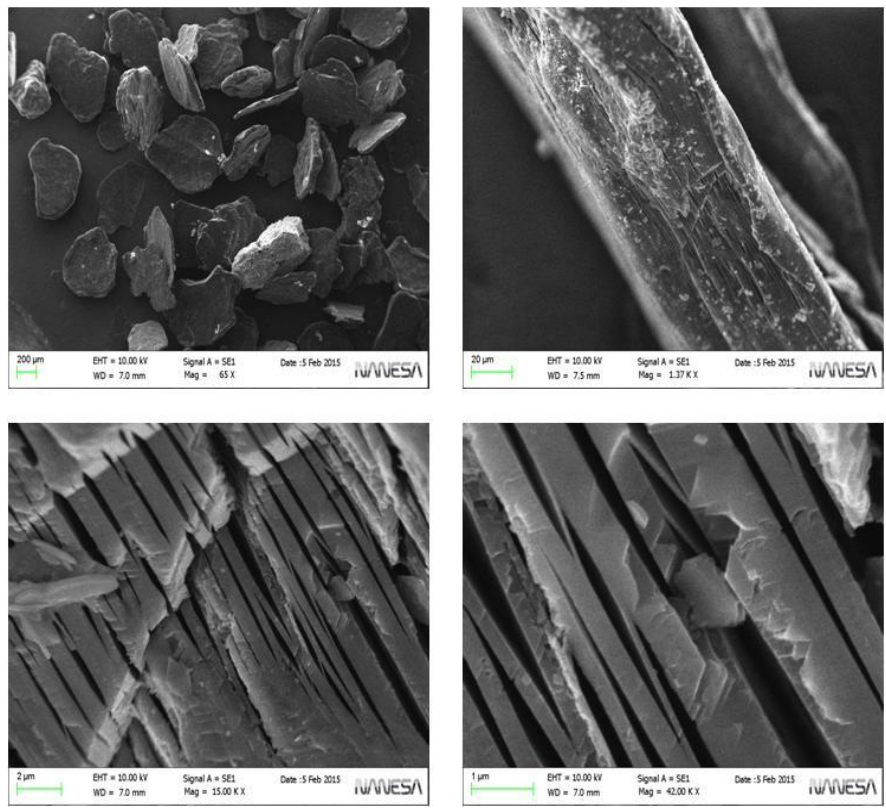

***Figure 11.*** *SEM images, Sample NNSa IG5 at various magnifications.*

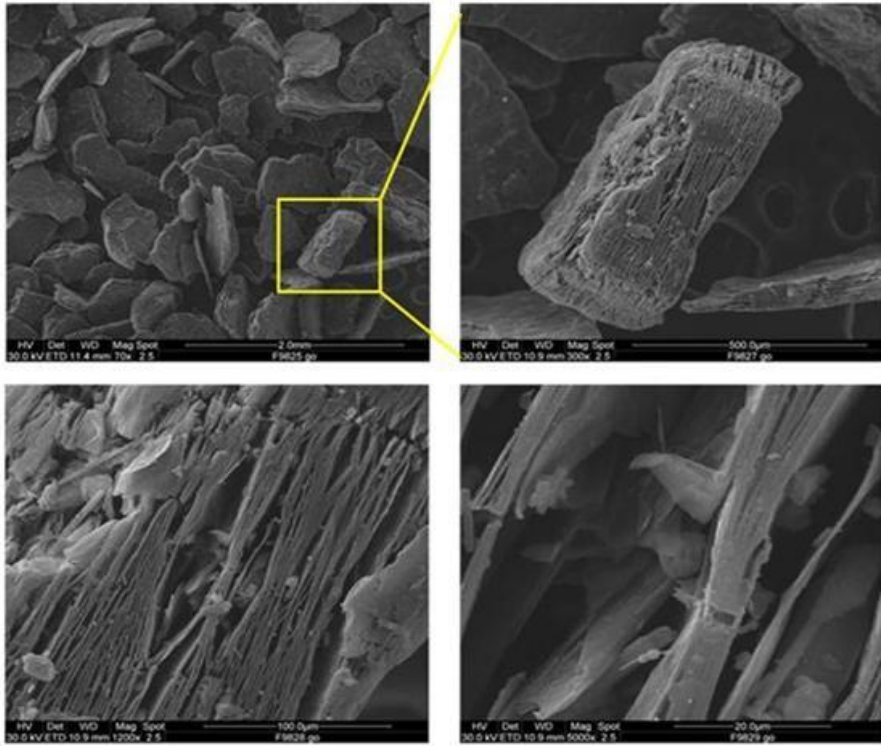

*Figure 12.* SEM images, Sample NNSa IG6 at various magnifications.

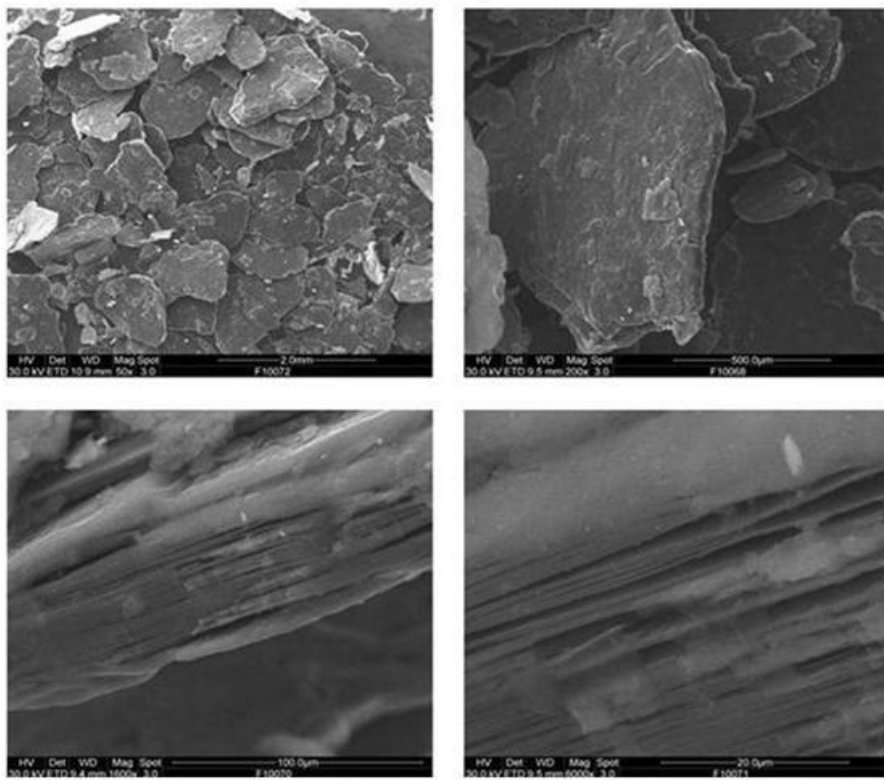

*Figure 13.* SEM images, Sample NNSa IG7 at various magnifications.

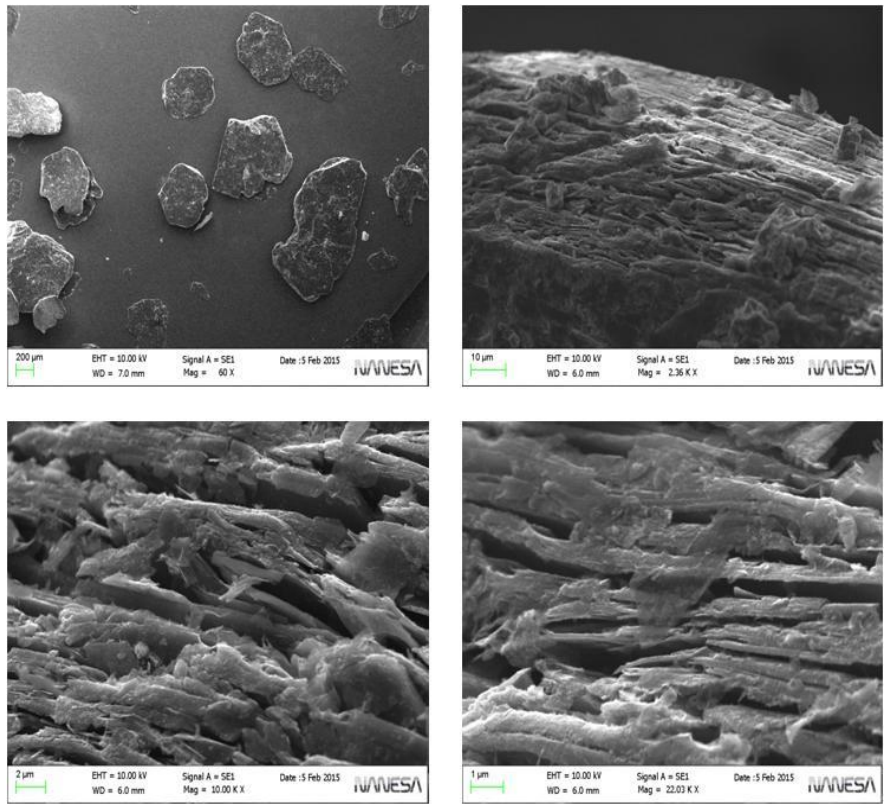

*Figure 14.* SEM images, Sample NNSa IG11 at various magnifications.

The images of expandable samples (IG samples) show the characteristic layered crystal structure of graphite. When viewed in cross-section these structures exhibit a partial separation due to the process of intercalation. This separation is more pronounced for the samples NNSa IG6 and NNSA IG11 than the others. The size of the flakes seems to correspond to the values declared by the suppliers, although in some cases dimension are larger. The sample NNSa IG2 is smaller than the other, indeed it is a micro-expandable graphite, which is confirmed by SEM images too. It was not possible to observe flake cross-section as they appeared covered by a layer of graphite. This effect is probably due to a mechanical process to make it smaller in size. Probably in this case the graphite was first intercalated and then subsequently milled to have a lower grain size.

### *2.4.2. Other characterisation of precursors*

### <u>*TGA*</u>

TGA were used to observe physical and chemical changes as a function of temperature. On the samples of GIC's this analysis was useful to study the thermal decomposition and vaporization of intercalating agents, in order to investigate on the thermal behaviour during the expansion process. The tests were performed from room temperature (20 °C) up to 1000 ° C using a heating ramp of 5 ° C / min., in air. Here reported two comparison Thermogram (Fig. 15). In the first one is compared the Weight loss of 4 precursors, heating ramp was 10 °C. In the second thermogram the comparison is between 2 precursors, heating ramp was 5 °C.

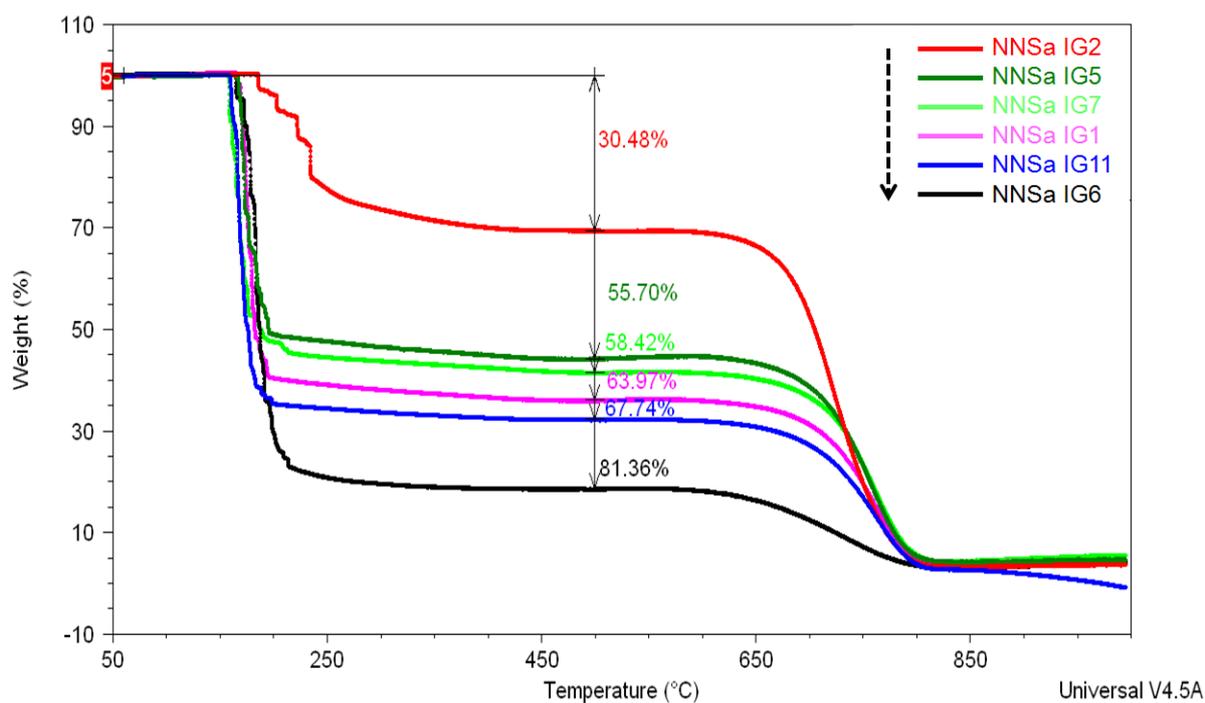

*Figure 15. Thermogram comparison of the different precursors*

*Table 2. Weight loss and residue percentage content of precursors*

| ID | First Weight Loss (%) | Second Weight Loss (%) | Residue (%) |
|---|---|---|---|
| NNSa IG1 | 63,9 | 31,4 | 4,7 |
| NNSa IG2 | 30,5 | 65,9 | 3,6 |
| NNSa IG5 | 55,7 | 38,9 | 5,3 |
| NNSa IG6 | 81,4 | 14,7 | 3,9 |
| NNSa IG7 | 58,4 | 36,4 | 5,2 |
| NNSa IG11 | 67,7 | 31,8 | 0,5 |

In all samples analyzed (Table 2), the first weight loss occurs at approximately 180-210 °C, it corresponds to degradation and evaporation of the intercalating agents. This is the critical range to consider in the process of thermal expansion. In the Table 1 are listed all values related to the weight losses. IG6 and IG11 are the precursors with the higher weight loss due to expansion (81,4% and 67,7%). The second weight loss occurs at approximately 600-800 °C, due to degradation of the carbon. The residue at 900 ° C is mainly composed of ash and impurities such as silica. The violent volume change during expansion, prevents an accurate measurement of the loss of weight. The results obtained from TGA can be affected by errors, overestimating the real weight loss. However, they are indicative to study the kinetics of degradation / evaporation of intercalating agents and obtain an estimate on the real degree of expansion.

## *XRF*

XRF analyzes were performed to analyze the elemental composition of the precursors. Due to the process of intercalation chemistry, they might contain residues of processing or extraneous elements (impurities). XRF analysis, however, does not allow identifying the light elements such as carbon, nitrogen and oxygen, then this is not a quantitative analysis. The precursors (expandable graphite) are carbon based material and contains in part also Oxygen, which is formed during the process of intercalation (see FT-IR analysis). Fluorescence Spectrophotometer allows to know the relationship between the percentages of heavy elements (such as Fe, Mn, Al, S). The amount of sulphur is an important parameter because it can be related to expansion rate. The sulphur originates from the sulphuric acid, one of the intercalating agent used for the process of intercalation of the graphite and the blowing agent that allows the transformation of expandable graphite to expanded graphite. In Fig. 16 are reported pie charts of six precursors. It shows the percentage ratio among the heavy elements for each precursor. The results obtained show that for almost all samples prevails sulfur compared to the other elements. The sulfur was part of the sulfuric acid which is the intercalating agent. So it is not correct to compare this to an impurity, also because almost all of it goes away in the form of gas during the expansion process. Also the manganese is present in significant quantities in some of them. The impurities found are the following elements: Fe, Si, Al, Na, K, Cr, Ca, Cl. Assuming that the sulfur and manganese are not impurities (this is not entirely true for manganese) but they are intercalating agents (I.A.). It was easily obtained the ratio among them, following:

*-NNSa IG1= 62% I.A. (38% impurities)*
*-NNSa IG2= 66% I.A. (34% impurities)*
*-NNSa IG5= 76% I.A. (24% impurities)*
*-NNSa IG6= 76% I.A. (24% impurities)*
*-NNSa IG7= 78% I.A. (22% impurities)*
*-NNSa IG11= 73% I.A. (27% impurities)*

These values are not the real content of impurities, but only a ratio among impurity content and intercalating agents.

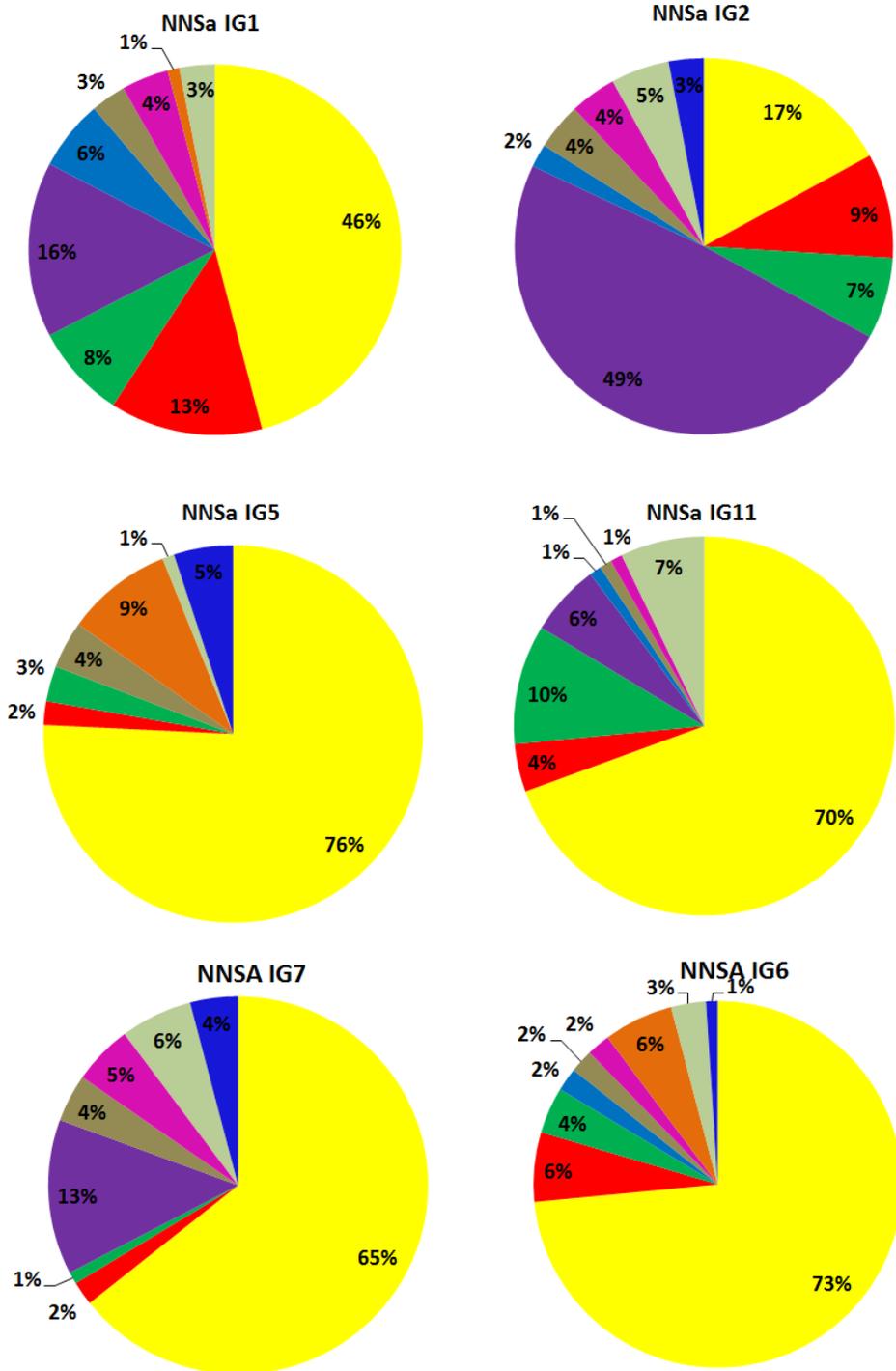

***Figure 16.*** *Pie charts of percentage ratio among heavy elements for each precursor.*

## FT-IR

The FT-IR analysis has been used for the identification of the functional groups formed, eventually, during the intercalation process (Fig. 17).

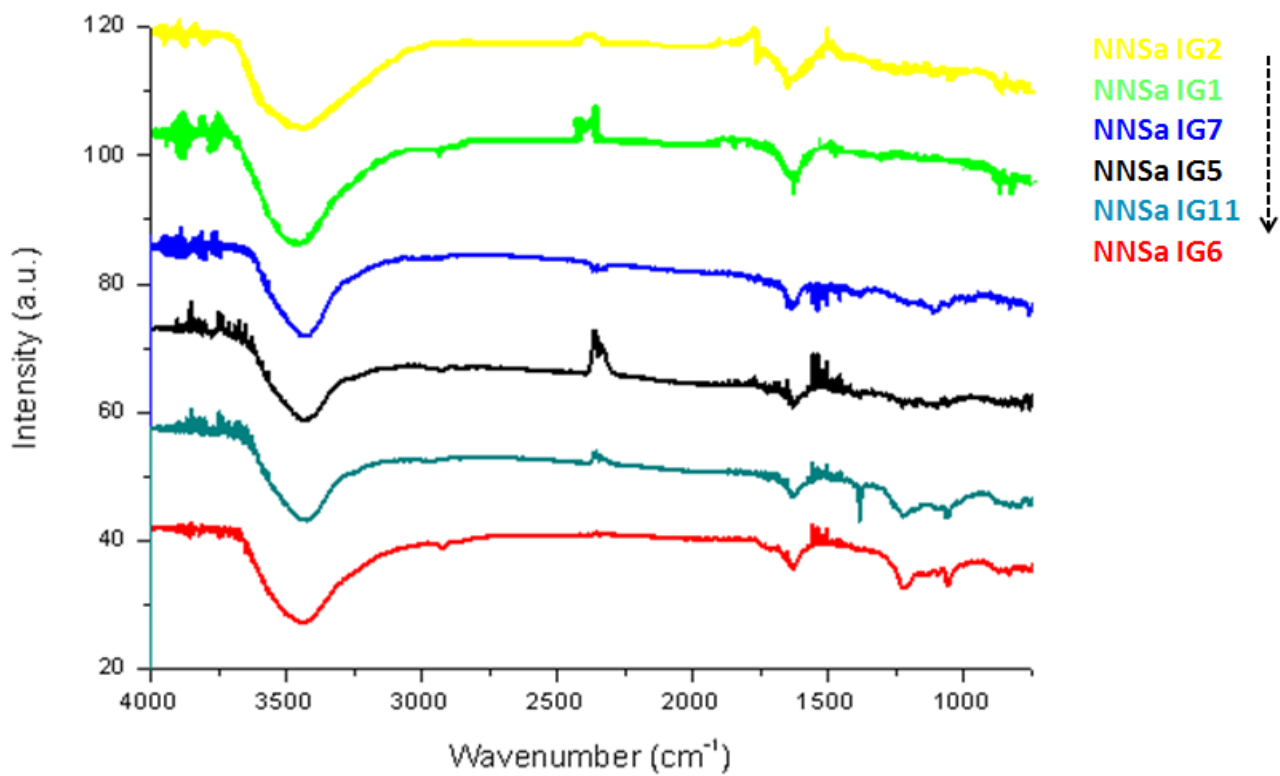

*Figure 17.* *Graph comparison of precursors analysed with FT-IR*

The samples show peaks very similar each other. The peaks in 1050 cm$^{-1}$ is due to stretching of S=O bond , 1228 and 1655 cm$^{-1}$ are due to stretching of CO and C = C bonds; the peak at 3400 cm$^{-1}$ instead due to stretching of the OH bond. The presence of CO bonds and OH indicate a partial oxidation of the graphite due to the process of intercalation with sulphuric acid and nitric acid.

## 3. RESULTS AND DISCUSSIONS

The precursors selected for the exfoliation tests have been well characterised to evaluate morphological and chemical properties. **SEM** images (Figs. 9-14) show the typical layered crystal structure of the graphite. The size of the flakes seems to correspond to the values declared by the suppliers, although in some cases dimension are larger. When viewed in cross-section these structures exhibit a partial separation due to the process of intercalation. This separation is more pronounced for the precursors NNSa IG6 and NNSa IG11. The sample NNSa IG2 is the smallest, indeed it is a micro-expandable graphite, and this is also confirmed by SEM images. In this case the flake cross-section appeared covered by a layer of graphite, losing the typical layered structure. This effect was probably due to a grinding process used by the supplier to make them smaller in size.

In **TGA** Analysis, Fig. 15, the first weight loss that occurs from 150 to 250 °C is related to vaporisation of intercalated agents, consequently it provides a rough estimation of intercalated agents quantity. Precursors NNSa IG6 and NNsa IG11 have the higher weight loss, 81.4% and 67.7% respectively. On the other hand the precursor NNSa IG2 has the lowest weight loss. According to these results, the precursors NNSa IG6 and NNSa IG11 are the products with the higher potential expansion rate. These values, in addition, confirm the first SEM observations that indicated higher separations among layers in their layered structures.

**XRF** results indicate the presence of heavy elements, in small amounts, in all precursors. Although these values are not the real content of heavy elements, pie charts (Fig. 16) can give a correct vision of ratio between impurities and heavy elements. The precursor with the best ratio, so with low impurities content, was NNSa IG11. This result confirmed the high purity declared by supplier.

Finally, the **FT-IR** analysis also confirm the best potential in terms of expansion rate for precursors NNSA IG6 and NNSA IG11. Only for them were observed two peaks related to the stretching of S=O and CO bonds, index of a partial oxidation of graphite and a higher degree of intercalation.

### 3.1. Lab process

Here have been reported results of characterisations performed on GNPs (graphite nanoplatelets) obtained downstream the process for expansion and exfoliation of the six precursors. The process used was on lab-scale, it has been already discussed in previous paragraphs.

### *PSA*

Particle size analysis was performed on each graphite nanoplatelets obtained by six precursors with lab-scale process. The distribution was typically Gaussian for all samples (Fig. 18).

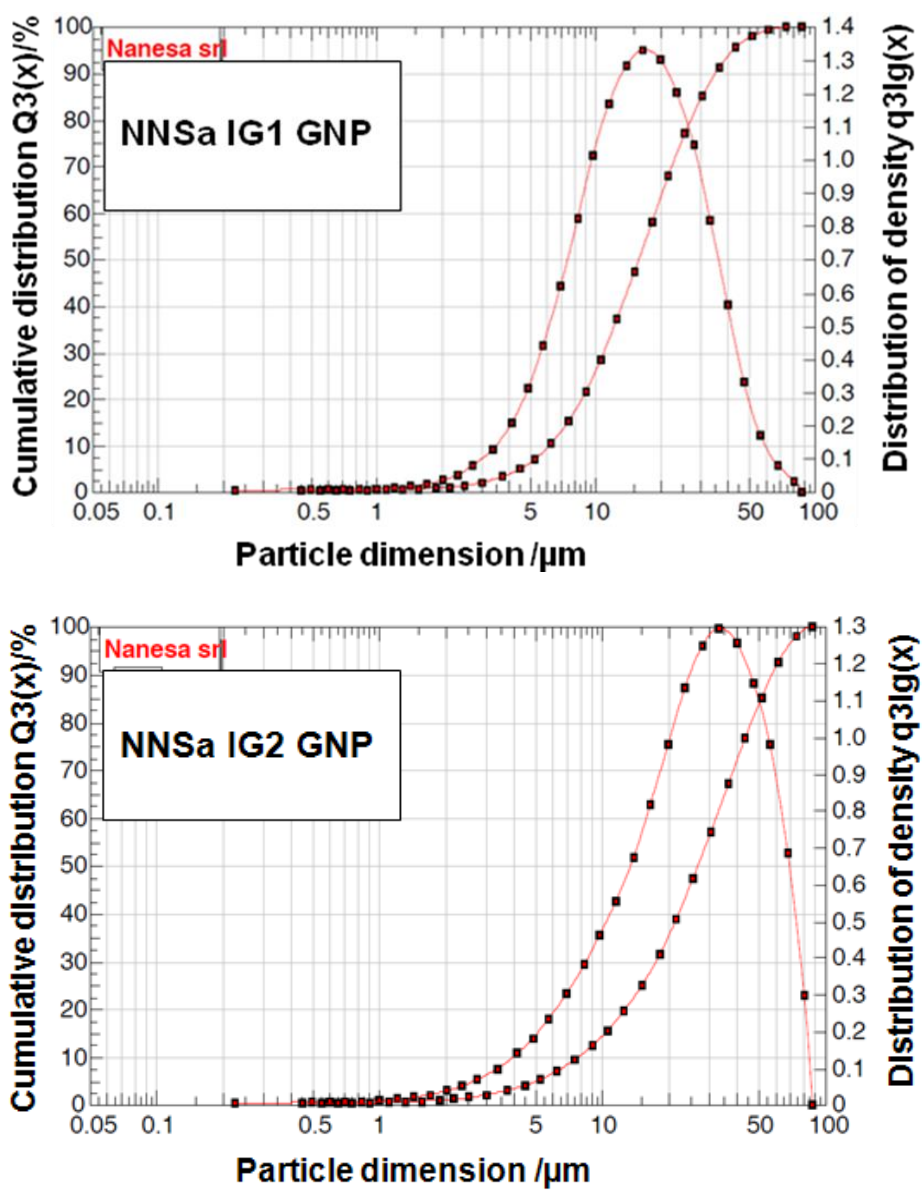

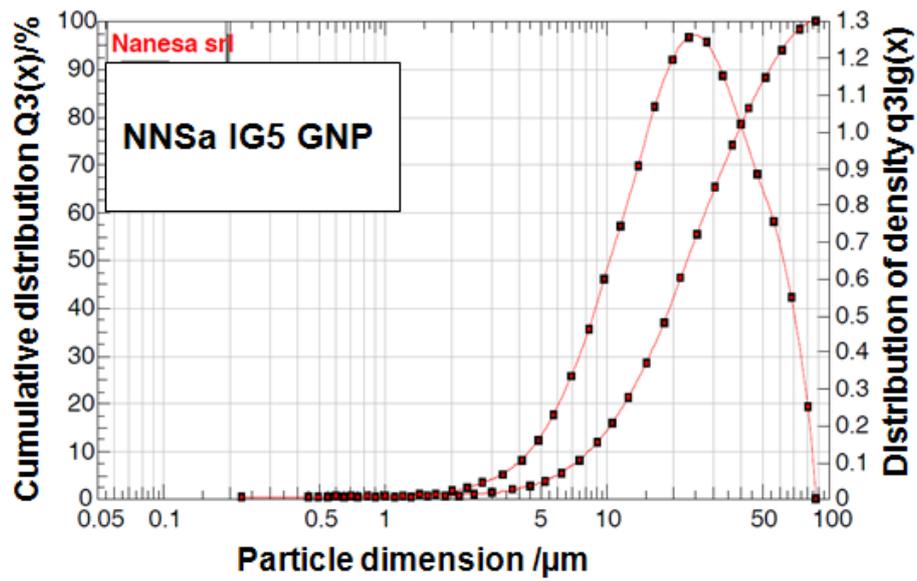
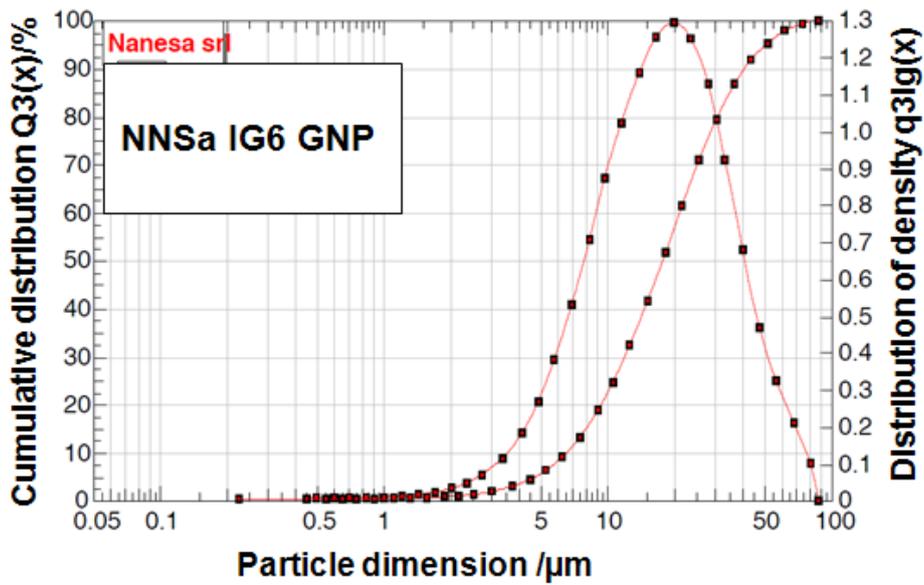
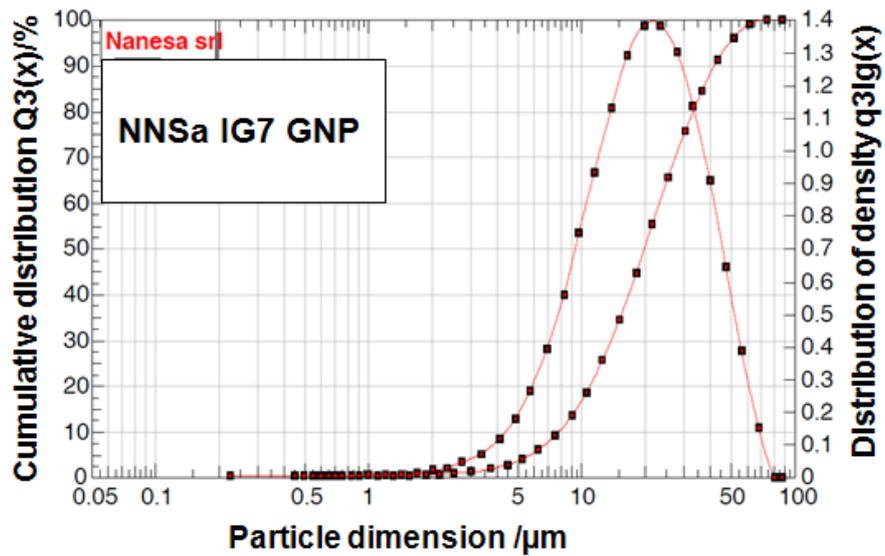

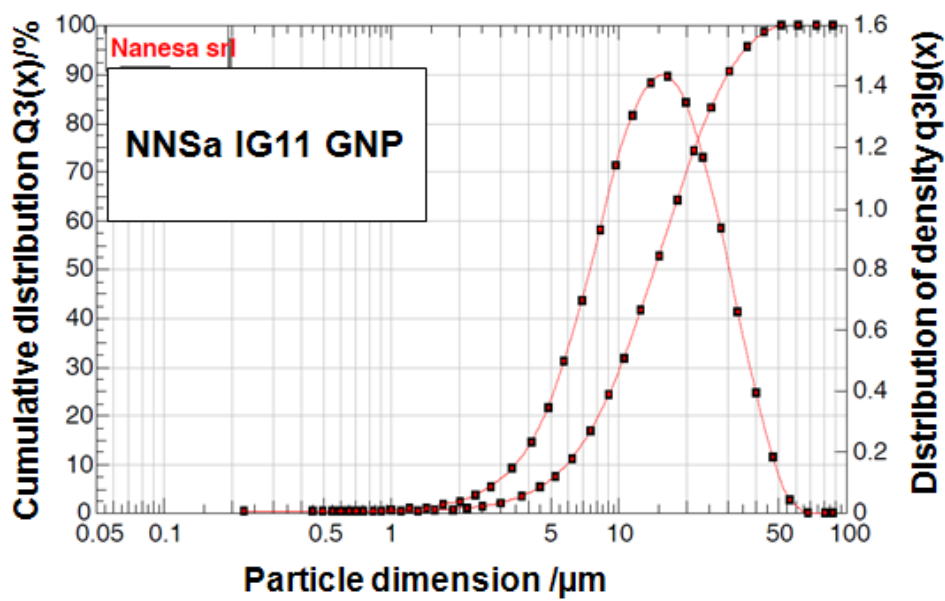

*Figure 18. The size distribution curves of each NNSa GNP obtained by lab-scale process*

In Table 3, the values of parameters obtained elaborating the individual curves are reported. The D10, D50 and D90 are commonly used to represent the midpoint and range of the particle sizes of a given sample. SMD is defined as the diameter of a sphere that has the same volume/surface area ratio as a particle of interest. VMD is the diameter of a hypothetical particle having the same averaged volume as that of the given sample. Assuming that thickness is negligible, In this case, VMD can be considered as a good estimation of particle lateral size.

*Table 3. Particle size values of NNSa GNPs*

| ID | D10 (µm) | D50 (µm) | D90 (µm) | SMD (µm) | VMD (µm) |
|---|---|---|---|---|---|
| NNSa IG1 GNP | 6.2 | 15.9 | 35.6 | 11.11 | **18.8** |
| NNSa IG2 GNP | 7.9 | 27 | 58.2 | 14.8 | 30.3 |
| NNSa IG5 GNP | 6.6 | 17.6 | 40.9 | 12.2 | 21.2 |
| NNSa IG6 GNP | 6.6 | 17.5 | 40.9 | 12.2 | 21.2 |
| NNSa IG7 GNP | 7.9 | 20 | 43.3 | 14 | 23.2 |
| NNSa IG11 GNP | 6 | 14.4 | 30.4 | 10.7 | **18.8** |

The sample NNSa IG2 GNP presents parameter values D90, D50 and D10 greater than the other samples. Although the pristine graphite of this sample was a micro-graphite (NNSa IG2) the nanoparticles obtained after exfoliation have the highest granulometry. This is fairly due to the fact that this graphite was actually a milled product and then the base unit (the nanoplatelets) is not smaller than the other, in terms of surface dimensions.The other samples show values more or less similar. Samples IG11 exf. and IG1 exf. have lower values than the other. is therefore conceivable that from these graphite may be obtained nanoplatelets with smaller surface dimensions.

### *SEM*

All samples of graphite nanoparticles obtained with lab-scale process have been studied also with SEM analysis (Figs. 19-24). For the sample preparation, each sample was dispersed in ethanol by Sonication, a drop of dispersion was deposited on a copper foil and dried in air to evaporate the solvent. Below have been reported micrographs by SEM. In a second step on each sample was performed EDS analysis (Energy Dispersive X-ray Spectrometry) with INCA software, in order to identify the constituent elements of the materials analysed.

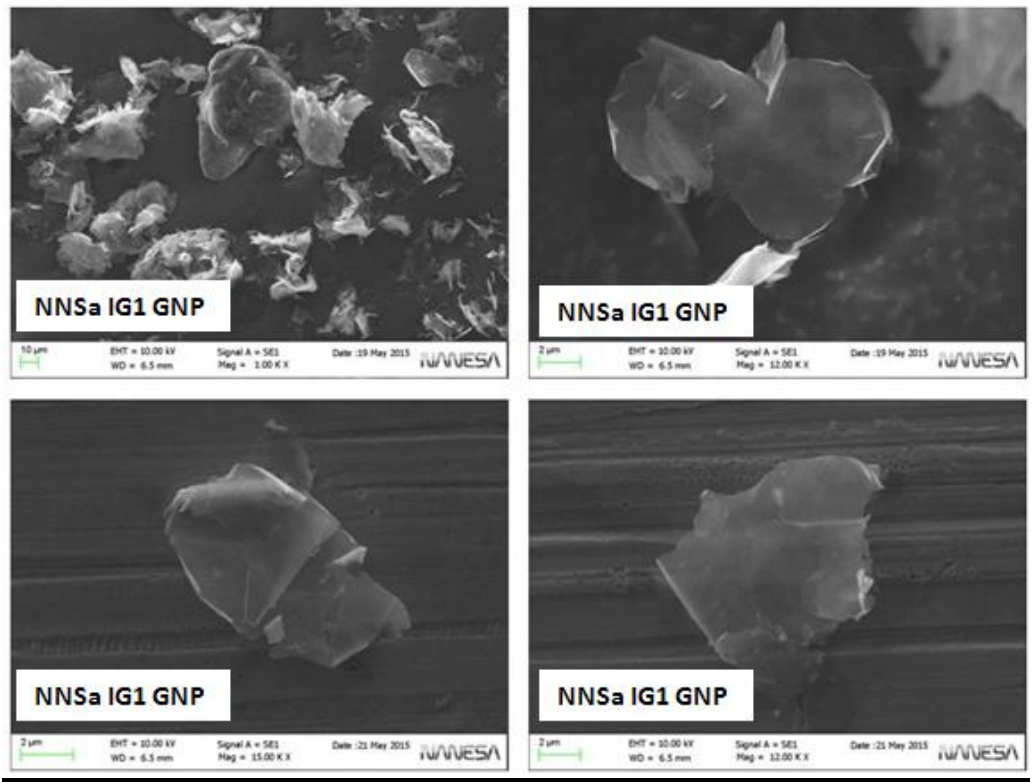

*Figure 19.* SEM micrographs of NNSa IG1 GNP. Overall the sample include some graphitic macrostructures, the isolated particles seem to have a good flatness, but far to be transparent (thicker). The edges of the particles are quite regular. The lateral size of the isolated particles was about 15 μm.

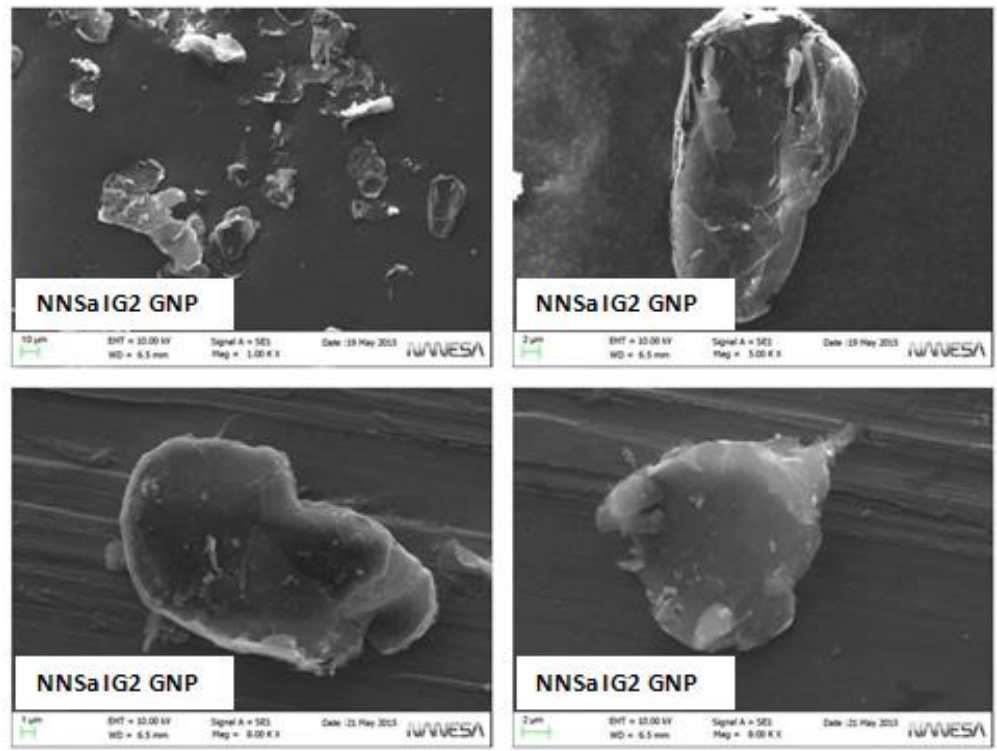

*Figure 20.* SEM micrographs of NNSa IG2 GNP. Overall the sample include many graphitic regions. It seems not to be well exfoliated. The particles are very regular and have smooth edges. They are well separated from each other, but are very thick structures (graphite). The lateral size of the isolated particles is around 20-30 µm.

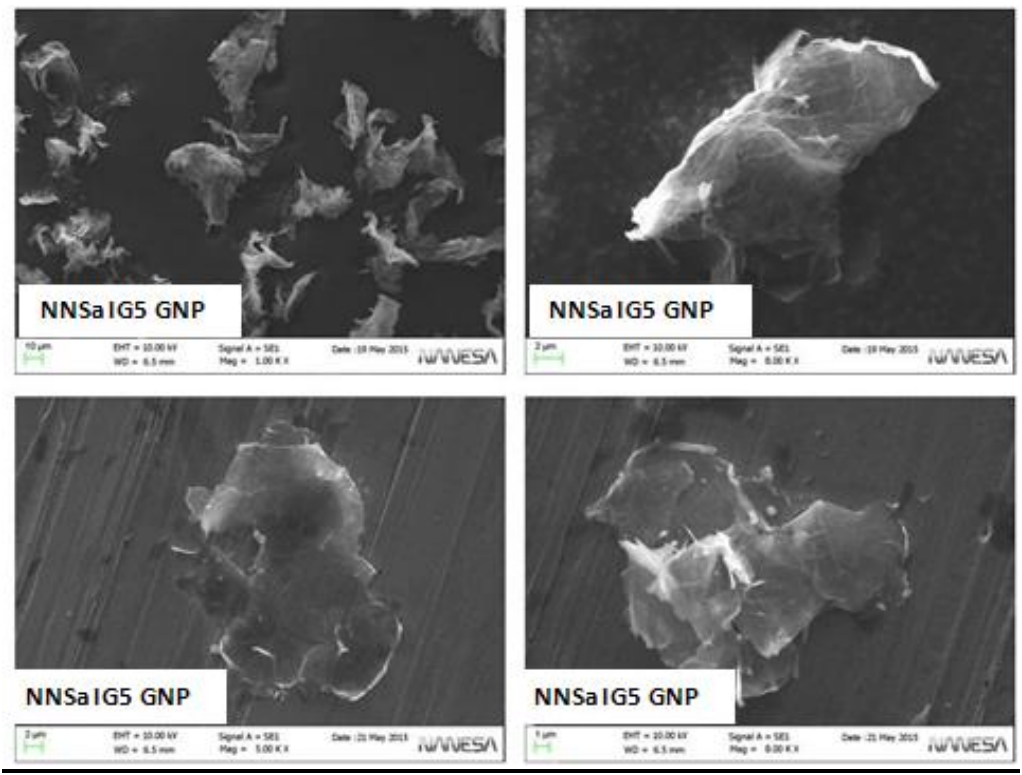

*Figure 21.* SEM micrographs of NNSa IG5 GNP. Overall the particles are very irregular structures and in many cases are observed fairly large agglomerates. The isolated particles have good transparency (low thickness) but have very irregular edges. The lateral size of the isolated particles is around 20 µm.

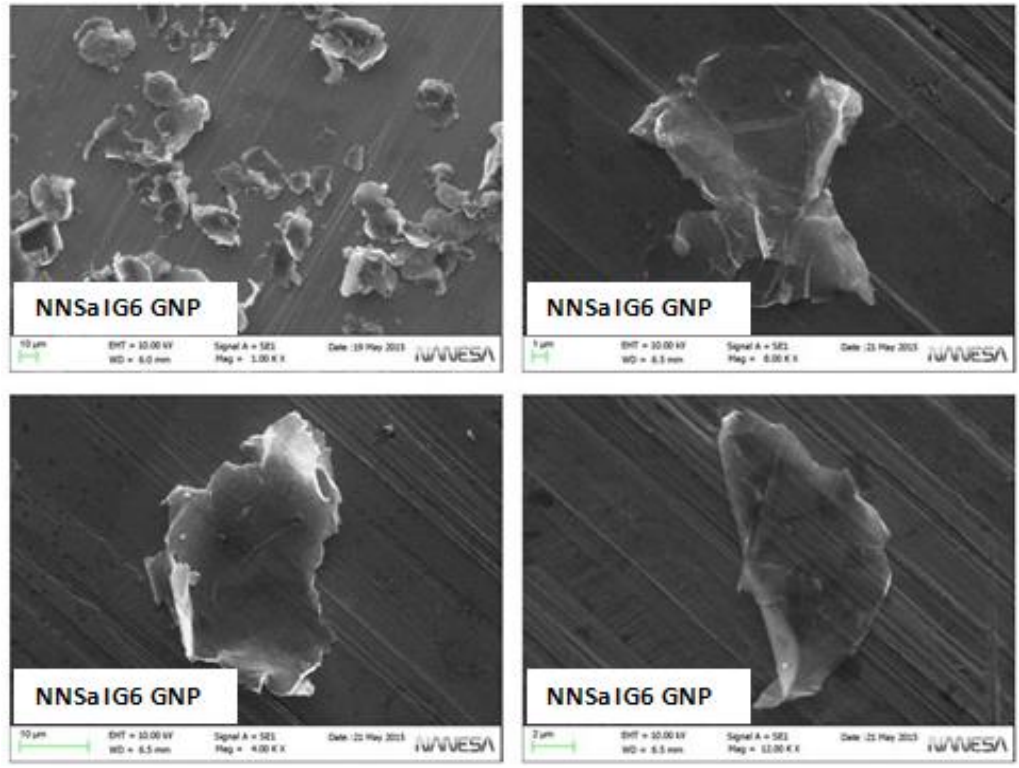

*Figure 22.* SEM micrographs of NNSa IG6 GNP. The particles as a whole are well dispersed and there are few agglomerates. The isolated particles have good flatness, high transparency (low thickness) and few irregular edges, lateral size is around 15-30 µm.

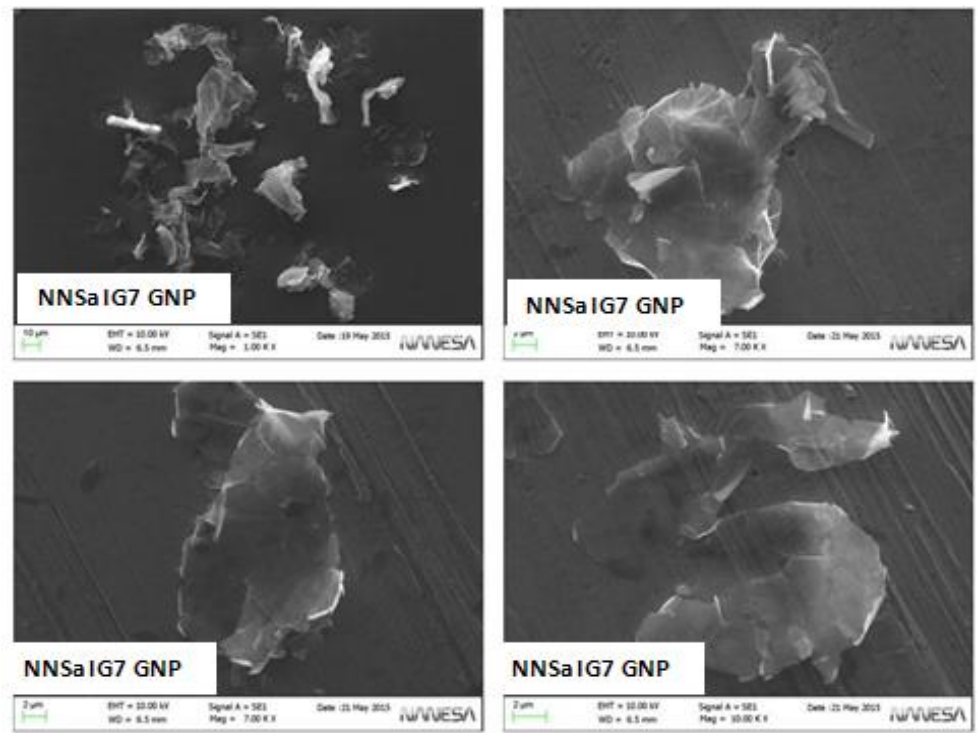

*Figure 23.* SEM micrographs of NNSa IG7 GNP. Also for IG7 ex, are noticed different graphitic agglomerated structures and particles have quite uneven edges, although there is a good transparency and flatness. The lateral size of the isolated particles is around 20-30 µm

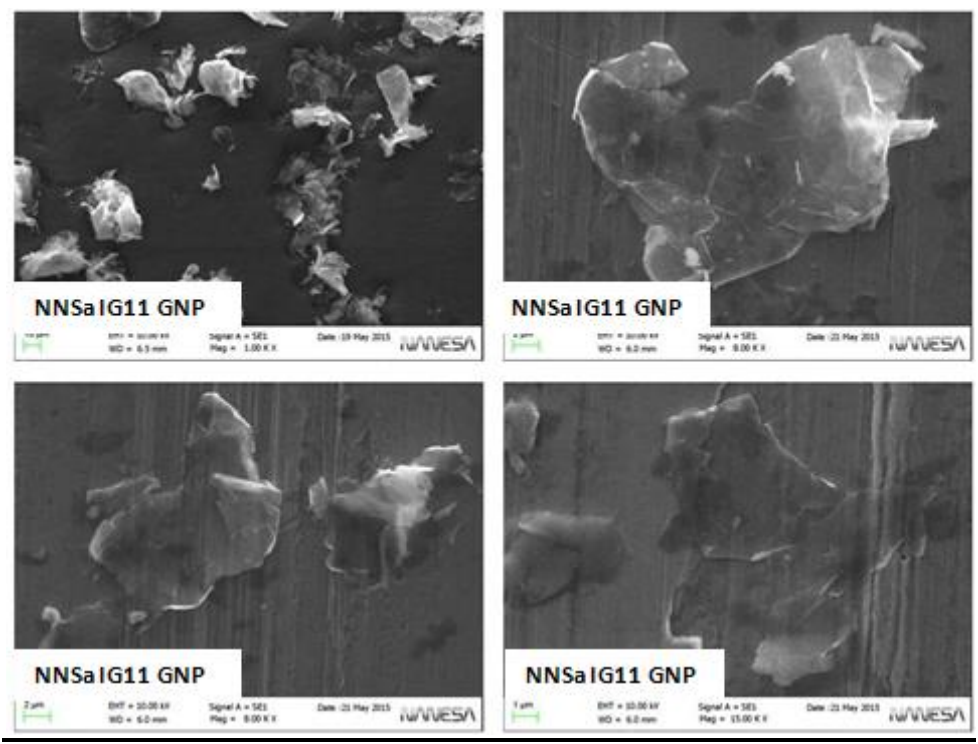

*Figure 24.* SEM micrographs of NNSa IG7 GNP. *Overall, even in this case are observable some agglomerates. The peculiarity of the isolated particles is that they seem to be very thin. They are very transparent and the edges are quite regular and planar. The lateral size of the isolated particles changes from 5 up to 20 µm.*

The SEM analysis are not quantitative analysis, but the images were useful to investigate the real morphology of exfoliated samples. These images, in fact, ensure that the product obtained with the precursor NNSa IG2 is not well exfoliated because it is formed by very thick micrometric particles. These results confirm those obtained with the TGA analysis of the precursors. The weight loss of the precursor NNSa IG2 appeared to be the lowest, this implied that this precursor had a lower content of intercalating agents and consequently the exfoliation was less effective than the other materials. The rest of samples show few differences among them. In some there are still traces of exfoliated graphite agglomerates (NNSa IG1 GNP and NNSa IG5 GNP). Samples NNSa IG6 GNP and NNSa IG11 GNP have the best features in terms of flatness, regularity of edges and thickness.

## EDS

On the same samples used for SEM characterisation was performed also EDS analysis (Energy Dispersive X-ray Spectrometry) in order to identify the constituent elements of the materials analysed. The EDS analysis was conducted using a copper foil as background material. It was used SEM in Back Scattering mode to identify the elements in the scanned areas (Fig. 25).

The Table 4 shows the values for the atomic weight percentage of the single elements found in detected areas. The EDS is qualitative and semi-quantitative analysis, then the values are to be considered as indicative. For all the samples obviously prevails the presence of Carbon (about 90%) and Oxygen (about 9%). In samples NNSA IG1 GNP and NNSa IG2 GNP are also visible traces of Sulphur, Silicon and Manganese. The high percentage of oxygen is due to the presence of this element also on the background (Cu foil), the EDS analysis of the background infact confirms the presence of this element (3.4%). Even the oxygen of silica (present as an impurity) and any remaining oxidizing agents could affect the total content of oxygen. In all samples analyzed it is possible to observe a quantity of carbon that reasonably exceeds 95%. Oxygen percentage was around 4-5 %, other elements are present only in the form of traces.

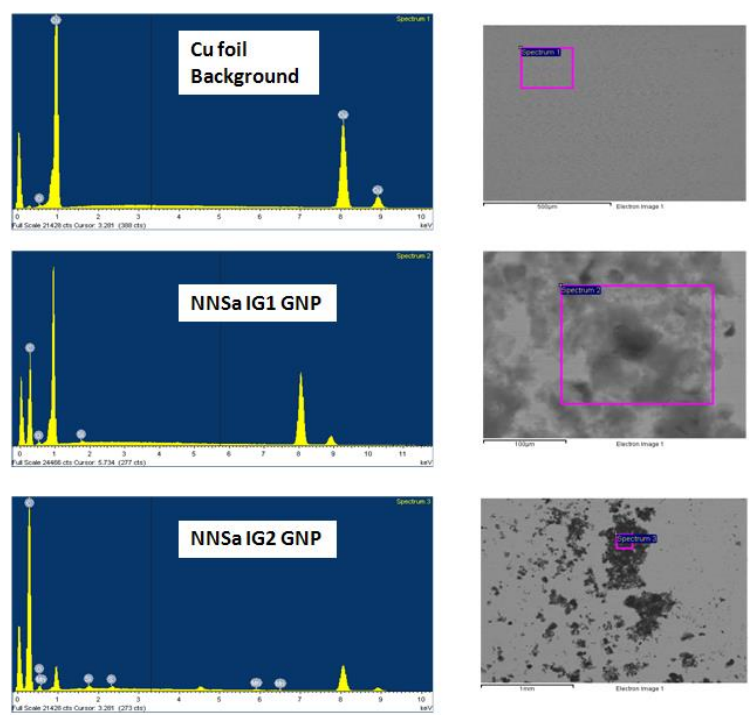

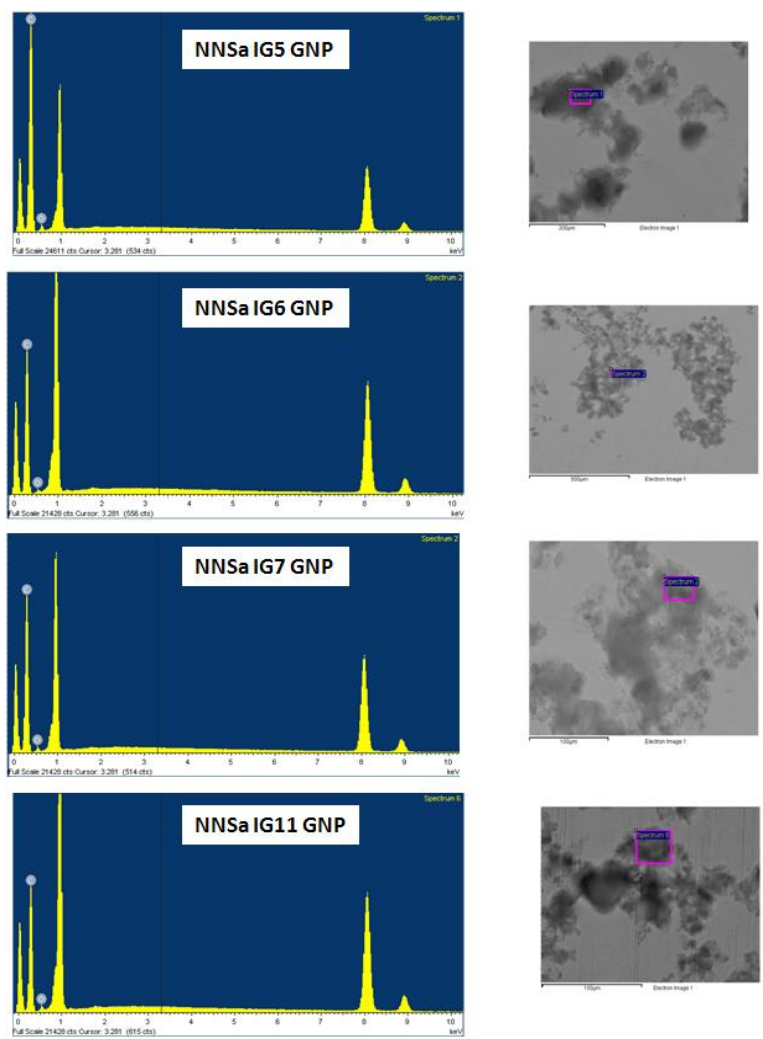

*Figure 25. Results for EDS analysis. On the left, the graphs with peaks relating to the signals of the individual elements. On the right, are shown viewing areas on which were carried out the analysis.*

*Table 4. Atomic weight percentage content of the single elements.*

| ID | C(%) | O (%) | Si (%) | Cu (%) | S (%) | Mn (%) |
|---|---|---|---|---|---|---|
| NNSa IG1 GNP | 90,4 | 9,3 | 0,24 | - | - | - |
| NNSa IG2 GNP | 91,7 | 7,8 | 0,21 | - | 0,17 | 0,14 |
| NNSa IG5 GNP | 93,3 | 6,4 | - | - | - | - |
| NNSa IG6 GNP | 91,6 | 8,4 | - | - | - | - |
| NNSa IG7 GNP | 91,1 | 8,9 | - | - | - | - |
| NNSa IG11 GNP | 89,2 | 10,2 | - | - | - | - |
| Background | - | 3,4 | - | 96,6 | - | - |

## XRD

The XRD analysis was performed on the exfoliated samples to obtain the diffraction pattern and information on the degree of stacking (Fig. 26).

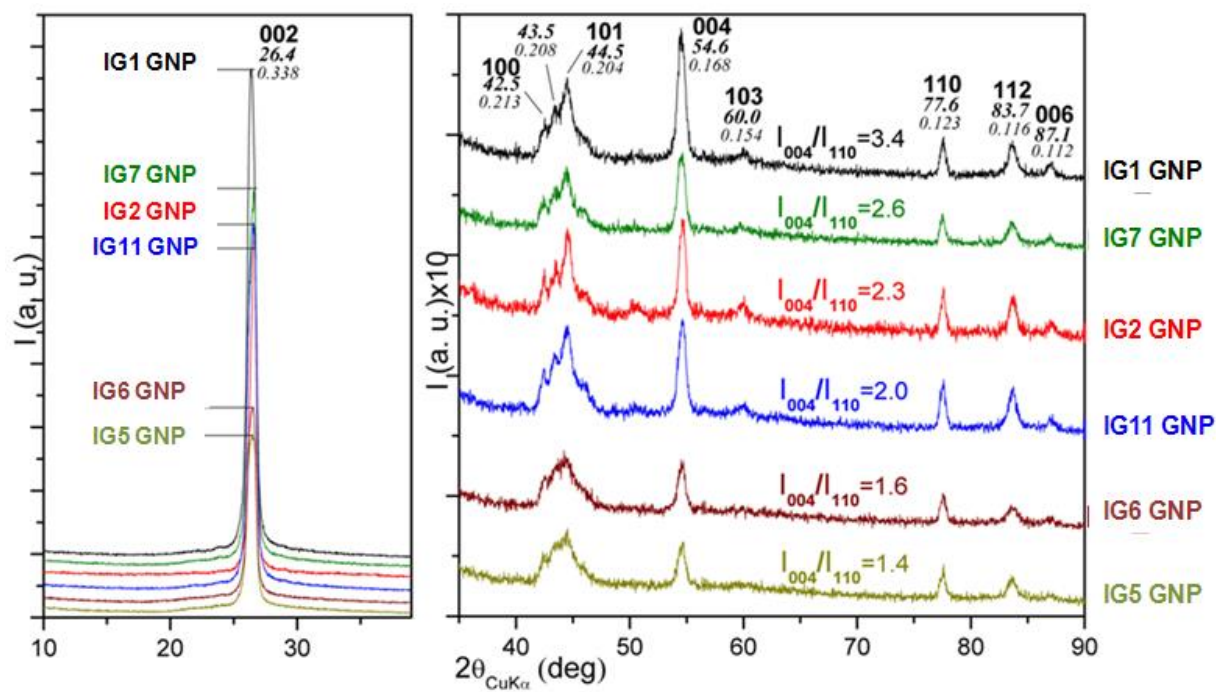

*Figure 26. X ray diffraction patterns for all samples*

The diffraction patterns are shown for all samples. The main diffraction peak is located at 2 θ = 26.4°, which correspond to the spacing between lattice planes 002 (Miller indices). This is the characteristic diffraction peak of graphite and compounds derived from graphite. The 002 peak of the graphite is a measure of interplanar spacing for two neighbouring graphene layers, corresponds to a d-spacing of about 0.34 nm, calculated with Bragg's law, Fig. 27, ($n\lambda=2d\ sin\theta$), where "θ" is the scattering angle, "λ" is X-ray wave length, "n" = 1 and "d" is d-spacing.

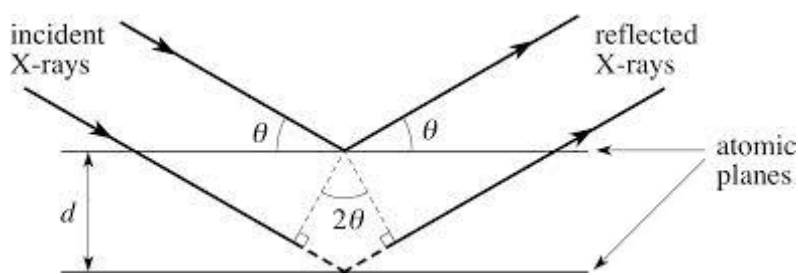

*Figure 27. Bragg's diffraction*

The ratio between the intensities of the peaks 004 and 110 indicates the degree of stacking of graphite, which decreases with this ratio. Samples IG5 ex, IG6 ex and IG11 ex have the lowest ratio, this indicates that the degree of stacking is smaller than the other. It is reasonable to assume that this is related to the characteristics of the initial precursors, which have a greater degree of intercalation.

Average crystallite size has been calculated from the line broadening of the diffractogram peaks using Scherrer formula (Eq.1). In the table X.xxx has been reported an estimate of the number of layers for each sample (Number of layers = D/d-spacing)

$$D = \frac{K\lambda}{B \cos \theta_B} \qquad (1)$$

*where*

D: Average Crystallite size; K: shape factor (0,89); $\lambda$: incident X-ray wavelength; $\beta$: the full width half maximum (FWHM) of diffraction peak expressed in radians; and $\theta$: peak position. Values of D and number of layers estimated with Scherrer formula are provide in Table 5.

*Table 5. Values of D and number of layers estimated with Scherrer formula*

|  | D (nm) | Layers |
|---|---|---|
| NNSa IG1 GNP | 11,08 | 33 |
| NNSa IG2 GNP | 13,30 | 39 |
| NNSa IG5 GNP | 9,98 | 29 |
| NNSa IG6 GNP | 10,78 | 32 |
| NNSa IG7 GNP | 10,64 | 31 |
| NNSa IG11 GNP | 10,15 | 30 |

Results are very similar, almost all samples have similar crystallite size. The sample NNSa IG2 has the higher D value and number of layers

### 3.2. Industrial process

Results of analysis performed on graphite nanoplatelets, processed on industrial plant, have been here reported. Precursors chosen and tested for industrial test were NNSa IG2, NNSa IG6, NNSa IG11. The reason for these choices are reported in Table 1.

### *PSA*

On dried samples of the industrial tests, were performed PSA analyzes to evaluate the particle size of nanoplatelets. The particle size distribution curves of the individual tests have been reported. They were compared with those obtained on the same precursors processed with the lab-scale method. From a comparison of the results obtained from PSA tests, laboratory versus Industrial tests (Fig.28), it was evident that only the product obtained from IG6 precursor has maintained the same characteristics as particle size of the laboratory tests (Table 6). The parameters D10 and D50 and VMD are slightly increased but results are good enough in terms of reproducibility. The samples NNSa IG2 GNP and NNSa IG11 GNP processed with the industrial plant, instead, had totally different characteristics from laboratory tests. The values D10, D50, D90 and VMD were, in both cases, much greater than those obtained by laboratory scale tests. Also the particle size distribution curves show different trends compared to those obtained from laboratory tests. In particular for the sample IG11 Ind process, the curve appears to have two different peaks. Most probably these two peaks are related to two different particle size distributions in the same product. This is reasonably due to a partial exfoliation of nanoparticles, some of them, those well exfoliated, have the correct size particle distribution, others are still present in aggregated form.

*Table 6. Particle size distribution parameters*

|  | D10 (µm) | D50 (µm) | D90 (µm) | VMD (µm) |
|---|---|---|---|---|
| NNSa IG6 GNP Lab. | 6.6 | 17.5 | 40.9 | 21.2 |
| NNSa IG6 GNP Ind. | 9.7 | 22.8 | 40.7 | 24.1 |
| NNSa IG2 GNP Lab. | 7.9 | 27 | 58.2 | 30.3 |
| NNSa IG2 GNP Ind. | 13.3 | 38.9 | 69.7 | 40.3 |
| NNSa IG11 GNP Lab. | 6 | 14.4 | 30.4 | 16.6 |
| NNSa IG11 GNP Ind. | 19 | 56.9 | 79.1 | 53 |

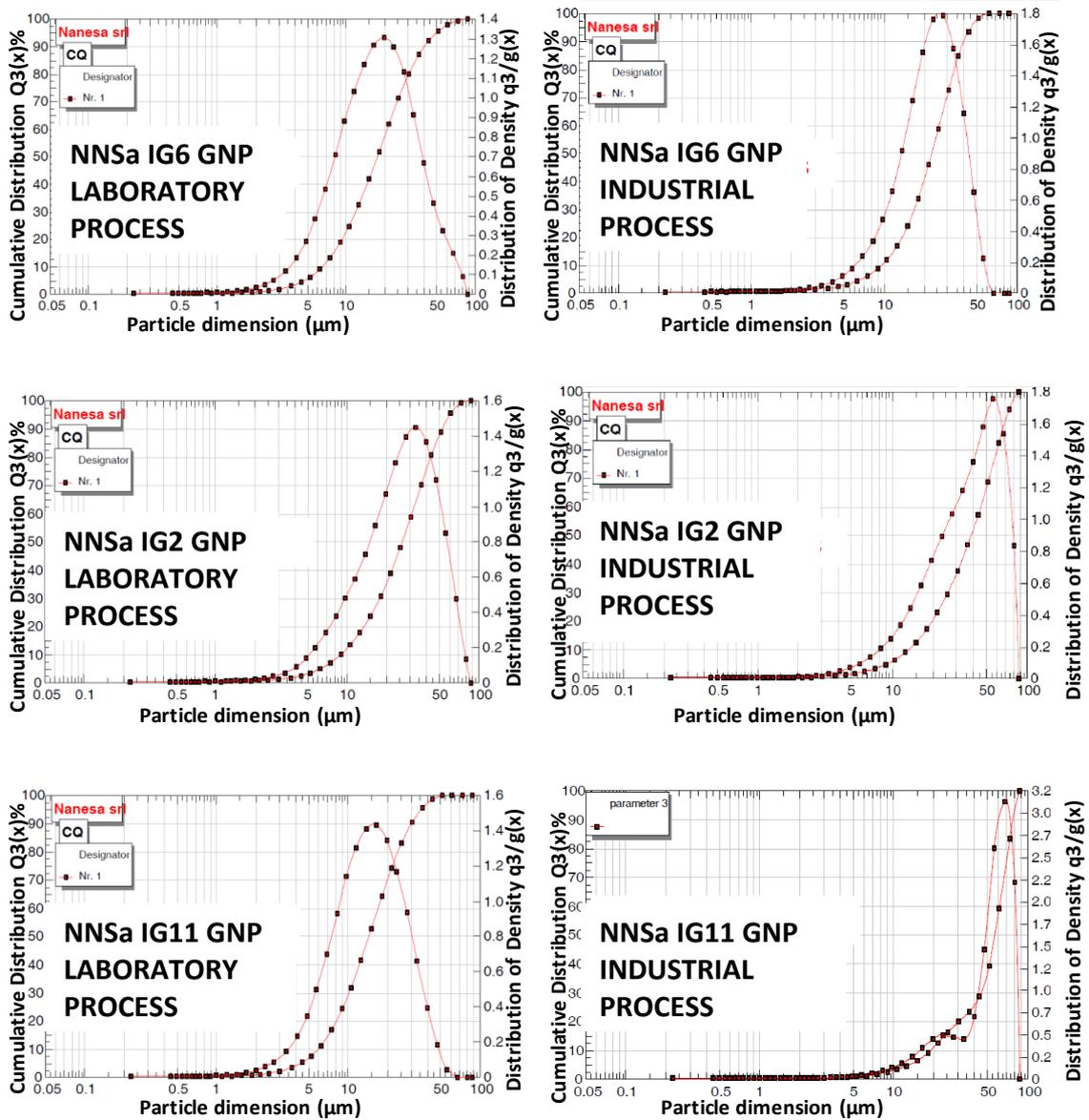

*Figure 28. Comparison of particle size distribution cruves (Lab vs Ind.)*

## OM

In Fig. 29, we reported some images obtained with optical Microscopy. The Magnifications were 100x and 200x. Samples analyzed were NNSa IG6 GNP, NNSa IG11 GNP, NNSa IG2 GNP (industrial process).

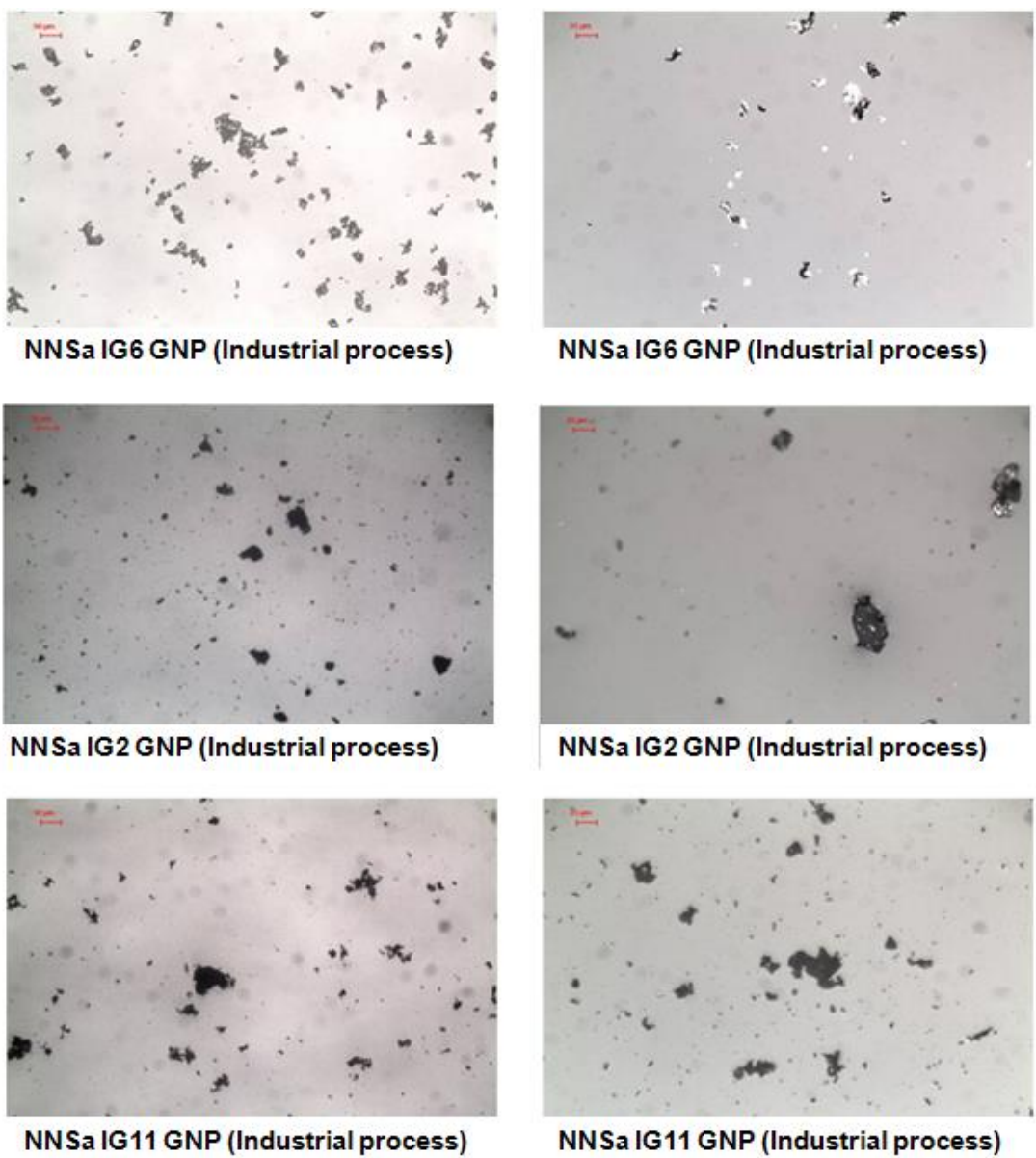

*Figure 29. Optical microscopy images. Magnifications used 100x (left) and 200x (right). Scale bar 50 μm (left), 25 μm (right).*

From these images was clear that for the samples NNSa IG2 GNP and NNSa IG11 GNP there are still many agglomerations. The NNSa IG6 GNP sample, instead, is composed of smaller particles, well distributed and more homogeneous.

## SEM

Here, (Fig. 30), we reported some images, obtained by SEM, of isolated particle. Samples analysed were NNSa IG6 GNP, NNSa IG11 GNP, NNSa IG2 GNP (industrial process).

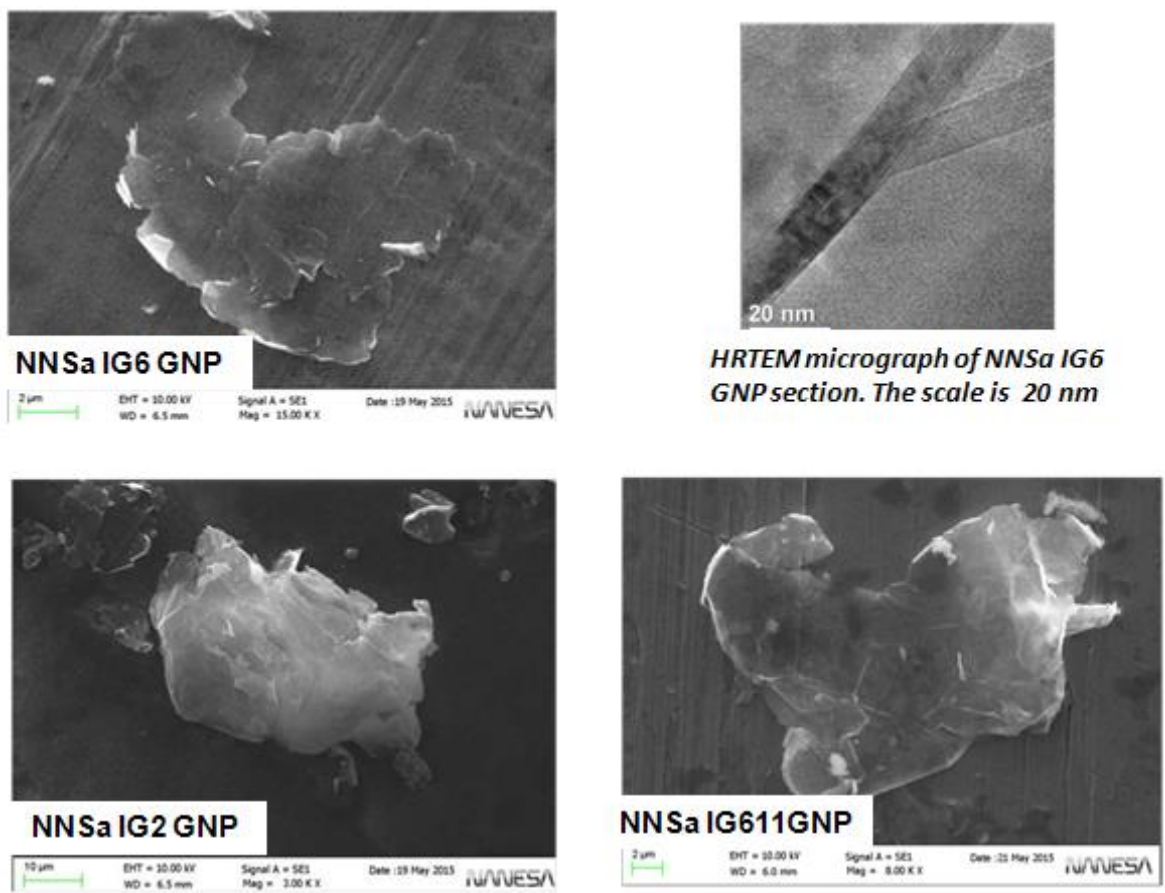

*Figure 30. SEM images for various NNSa samples*

SEM images obtained from the analysis confirm the results of the previous tests (PSA-OM). The product NNSa IG6 GNP has good flatness and transparency, it is very similar to that obtained in the laboratory test. On the other hand, samples NNSa IG2 GNP and NNSa IG11 GNP are very different when compared to the samples obtained in the laboratory, in particular the IG11 sample. It seems that the industrial process was not effective enough to exfoliate the nanoparticles with respect to the laboratory process. It is also reported the image of a graphene nanoplatelet (NNSa IG6 GNP) viewed in section. It was obtained with HR-TEM. The thickness of the single nanoplatelet measures 14 nm.

## 4. CONCLUSIONS

Based on laboratory results were selected some precursors to test on industrial pilot plant. The expandable graphite IG6, IG2 and IG11 have been tested on the continuous system plant. The operating conditions and the parameters of expansion and exfoliation have been set in order to obtain the best thermal expansion and a homogeneous and efficient exfoliation. In order to have a real comparison between the various precursors, all the plant parameters have been set in the same manner for the three processes. As regards the precursor IG6, already during the Sonication phase the product reached a good degree of exfoliation. This was suggested by the viscosity measurements performed during the process. The dispersion, in fact, varied from a viscosity of 10 cP, after 1 hour of treatment, up to 130 cP, after 6 hours of treatment. Morphological characterizations have confirmed that the product was entirely similar to that obtained by the lab-scale. The PSA analysis, SEM and OM confirm these results. As for the micro-graphite IG2 the results were different from expectations. Being an expandable micro-graphite was reasonable to expect a product smaller in size than the "usual" expandable graphite. The tests on the continuous plant, instead, confirmed the indications suggested previously with preliminary tests. The product, obtained from micro-graphite IG2, in fact, was less exfoliated and coarser if compared with the product obtained with the expandable graphite IG6. This is confirmed by all analysis. The PSA, XRD, SEM, OM, have confirmed that. Also the viscosity measurements carried out during the exfoliation process confirms that the degree of exfoliation achieved was not optimal. It has gone from 4 to 17 cP after 10 hours. The predominant reason for these results has to be found in the degree of intercalation of this graphite. Most likely this graphite is reduced in size with mechanical grinding processes that reduce the amount or the ability to absorb the intercalating agents. The result is a micro-graphite with a low degree of intercalation that affects all subsequent stages of the process. For Graphite IG11 the laboratory results were encouraging. From this graphite were obtained particles with the lowest lateral size. The analyses indicated that was reached a good degree of exfoliation. The test in the industrial plant, however, did not give good results. The PSA analysis suggests that there were still large graphite agglomerations.

**ACKNOWLEDGE**

The research leading to these results has received funding from the European Union Seventh Framework Programme under grant agreement No. 604391 and Horizon 2020 Programme under grant agreement No. 696656 Graphene Flagship

**DATA AVAILABILITY**

The raw/processed data required to reproduce these findings cannot be shared at this time as the data also forms part of an ongoing study.